# The response of amino acid frequencies to directional mutation pressure in mitochondrial genome sequences is related to the physical properties of the amino acids and to the structure of the genetic code.

Daniel Urbina[1], Bin Tang[1,2] and Paul G. Higgs[1].

1. Department of Physics and Astronomy, McMaster University,
Main Street West, Hamilton, Ontario L8S 4M1, Canada.

2. Division of Genomics and Proteomics, Ontario Cancer Institute,
University of Toronto, Suite 703, 620 University Avenue, Toronto, ON M5G 2M9, Canada

Corresponding author: Paul Higgs
email: higgsp@mcmaster.ca
Tel 905 525 9140 ext 26870;        Fax 905 546 1252

**Abstract**

The frequencies of A, C, G and T in mitochondrial DNA vary among species due to unequal rates of mutation between the bases. The frequencies of bases at four-fold degenerate sites respond directly to mutation pressure. At $1^{st}$ and $2^{nd}$ positions, selection reduces the degree of frequency variation. Using a simple evolutionary model, we show that $1^{st}$ position sites are less constrained by selection than $2^{nd}$ position sites, and therefore that the frequencies of bases at $1^{st}$ position are more responsive to mutation pressure than those at $2^{nd}$ position. We define a measure of distance between amino acids that is dependent on 8 measured physical properties, and a similarity measure that is the inverse of this distance. Columns 1, 2 3 and 4 of the genetic code correspond to codons with U, C, A and G in their $2^{nd}$ position, respectively. The similarity of amino acids in the four columns decreases systematically from column 1 to 2 to 3 to 4. We then show that the responsiveness of $1^{st}$ position bases to mutation pressure is dependent on the $2^{nd}$ position base, and follows the same decreasing trend through the four columns. Again, this shows the correlation between physical properties and responsiveness. We determine a proximity measure for each amino acid, which is the average similarity between an amino acid and all others that are accessible via single point mutations in the mitochondrial genetic code structure. We also define a responsiveness for each amino acid, which measures how rapidly an amino acid frequency changes as a result of mutation pressure acting on the base frequencies. We show that there is a strong correlation between responsiveness and proximity, and that both these quantities are also correlated with the mutability of amino acids estimated from the mtREV substitution rate matrix. We also consider the variation of base frequencies between strands and between genes on a strand. These trends are consistent with the patterns expected from analysis of the variation among genomes.

**Keywords**

## 1. Introduction

Complete mitochondrial genome sequences are now available for hundreds of metazoan species. Our own OGRe database has been set up for comparative analysis of these genomes (Jameson *et al.* 2003). With a few minor exceptions, the same set of 13 protein-coding genes is found on all these genomes. These sequences provide an ideal data set with which to study the influence of directional mutation pressure on the frequencies of bases in gene sequences and the corresponding variation in the frequencies of the amino acids they encode.

The term directional mutation pressure refers to situations in which the rates of forward and reverse mutations between the DNA bases are not equal, so that mutation drives the base frequencies away from the balanced state of 25% each. Previous studies have emphasized the role of directional mutation pressure in determining the G+C content of genomes. Sueoka (1988) showed that G+C content varies widely among genomes. Variation is greatest at the third codon position because many substitutions at third position sites are synonymous, whereas all second position substitutions and almost all first position substitutions are non-synonymous. When many genomes are compared, there is a clear correlation between the G+C frequencies at the first or second position and those at the third position. The slope of the regression line depends on the strength of the selective constraint acting on first and second position sites relative to third position sites. This work has been extended to a wide range of species (Sueoka, 1995, 1998). Directional mutation pressure on the G+C content of genomes appears to vary fairly rapidly among species, and is often strong enough to influence the frequencies of amino acids in the protein sequences coded by the genomes (Lobry 1997; Foster *et al.* 1997; Singer and Hickey 2000; Knight *et al.* 2001a,b, Bharanidharan *et al.* 2004). The fact that base frequencies at non-synonymous sites vary less than those at synonymous sites but follow the same trends leads to the conclusion that amino acid frequencies are changing in response to the change in G+C content rather than that selection on amino acid frequencies has driven a change in G+C content. Singer and Hickey (2000) found that the GARP amino acids, which have G or C at both first and second positions in their corresponding codons, show a clear increase in frequency with increasing G+C content, whereas the FYMINK amino acids, which have A or T at both first and second positions, show a clear decrease in frequency with increasing G+C.

All these studies consider G+C as the principal variable describing DNA base frequencies. The rules of complementary base pairing in double-stranded DNA imply that the frequency of G on one strand will be equal to the frequency of C on the other strand and similarly for A and T. If the two strands are equivalent in terms of the mutation and selection acting on them, then the frequency of any base on one strand should be equal to the frequency of the same base on the other strand (within statistical noise). In consequence, on each strand, the frequency of G must equal that of C and the frequency of A must equal that of T. This is termed Parity Rule 2 (or PR2 - Sueoka, 1995). When PR2 applies, there is only one degree of freedom for base frequency variation, and it is sufficient to describe a sequence by its G+C content only.

However, in mitochondrial genomes, the two strands are not equivalent, due to the asymmetry of the replication process, and PR2 does not apply. The asymmetry of the strands has been demonstrated specifically by looking at mutations in human mitochondrial DNA (Tanaka and Ozawa 1994) and is apparent in comparative studies across mitochondrial genomes of different species (Reyes *et al.* 1998; Knight *et al.* 2001b; Bielawski & Gold, 2002; Faith & Pollock, 2003; Krishnan *et al.* 2004). Mechanisms leading to asymmetry of the strands are discussed further in Section 8 below. However, the main point of this paper is to develop a model of directional mutation pressure that allows all four base frequencies to vary. Although violation of PR2 is not limited to mitochondria (see Sueoka 1999, and McLean *et al.* 1998 for examples in bacterial genomes), the effect is sufficiently strong in mitochondria that theories based on a single G+C variable are clearly inadequate, and a more complex theory is required.

Non-stationary base and amino acid frequencies are a potential source of bias in phylogenetic studies with mitochondrial sequences (Foster & Hickey, 1999; Schmitz *et al.* 2002; Bielawski & Gold 2002; Krishnan *et al.* 2004; Gibson *et al.* 2005; Raina *et al.* 2005), and this provides one motivation for the present study. More fundamentally, however, the large variation in amino acid frequencies arising from mutation pressure suggests that mitochondrial sequences (at least in some species) may be far from optimal because of the presence of a large number of deleterious amino acid changes. Amino acid frequencies respond to different extents to the mutation pressures on the bases. We will show that the responsiveness of an amino acid is influenced by the genetic code structure and by the physical properties of the amino acids. It is known from studies of amino acid substitution rate matrices such as PAM (Dayhoff *et al.* 1978; Jones *et al.* 1992) and mtREV (Adachi and Hasegawa, 1996) that changes between amino acids with similar properties are more frequent because they are less disruptive to protein structure and are less likely to be eliminated by stabilizing selection. Here we use a measure of amino acid similarity that depends on a set of eight experimentally measured physico-chemical properties. We show that the amino acids that vary most in response to directional mutation pressure are those whose neighbouring amino acids in the genetic code (those accessible by single mutations) are most similar.

## 2. Data and Notation

All the sequence data used in this paper are from publicly available completely sequenced mitochondrial genomes of metazoan species. These sequences have been incorporated into our own relational database, OGRe (Organellar Genome Retrieval), as described by Jameson *et al.* (2003). The latest version of OGRe is available on line at http://ogre.mcmaster.ca . The web site shows graphical information on the frequencies of bases, codons and amino acids in mitochondrial genomes, allows full codon usage tables to be downloaded and includes additional features for visualization of gene order. The present paper analyzes several different data sets downloaded from OGRe. The metazoan data set consists of all 473 genomes that were in OGRe in July 2004. Two subsets were also considered: the mammal set consists of 109 species, and the fish data set consists of 172 actinopterygians (ray-finned fish). The mammals and ray-finned fish form two comparable but independent monophyletic groups.

In all the metazoan species there is a preponderance of genes on the plus (or H) strand. In several groups all genes are on the plus strand (these include all known examples of Annelids, Brachiopods and Platyhelminthes, and most Nematodes). In vertebrates, 12 of the 13 protein-coding sequences are on the plus strand. For most of this paper we will therefore consider only the plus strand, although comparison between strands is discussed in Section 8.

We will use the notation N1, N2, and N3 for the frequency of base N (= A, C, G or U) at the three codon positions. We will use the notation N4 for the frequency of base N at fourfold degenerate (FFD) sites, *i.e.* in the third codon position of 4-codon families. There are eight 4-codon families with FFD third positions: CUN (Leu); GUN (Val); UCN (Ser); CCN (Pro); ACN (Thr); GCN (Ala); CGN (Arg); and GGN (Gly). The third codon positions of these codons are not subject to selection at the protein level and therefore they should respond rapidly to changes in mutation pressure.

Figure 1 shows the relationships between the N4 frequencies in the Metazoa data set. Figure 1(a) shows that there is a strong negative correlation between C4 and U4. U4 was chosen as the horizontal axis because it covers the widest range of all the bases - from 6.7% in the snake *Leptotyphlops dulcis* to 92.4% in the nematode *Strongyloides stercoralis*. When PR2 applies, A4 and U4 should be equal; however, Figure 1(b) shows that A4 is almost independent of U4 when U4 < 40% and decreases for higher values of U4. G4 is low in all species and shows no trend as a function of U4. Figure 1(c) shows that G4 tends to decrease with A4, and that A4 also covers a very wide frequency range. It is clear from these figures that PR2 does not apply, and that it is insufficient to describe the sequences merely by their G+C content.

## 3. Variation of base frequencies at 1st and 2nd positions

The wide range of N4 frequencies shows that there is a strong directional mutation pressure away from equal base frequencies, and that the magnitude and direction of the mutation pressure varies among species. We will suppose that the DNA sequences of a given species are evolving according to a mutational model in which the equilibrium frequency of base N is $\pi_N$. We will assume that the FFD sites rapidly reach equilibrium under mutation and hence that the observed values of N4 for a given species are direct indicators of the $\pi_N$ frequencies in that species. Although FFD sites are not influenced by selection on amino acids, other selective effects at the DNA level might apply. We will assume that the mutation rate is sufficiently large that any weak DNA-level selection is negligible (we return to this point in the discussion).

In contrast, the base frequencies at first and second positions are influenced both by selection on non-synonymous substitutions and by mutation pressure. Figure 2 shows the variation in frequencies of each of the four bases at positions 1 and 2 as a function of the frequency of the same base at the FFD sites. The full metazoa data set is shown. Figure 3 shows the same quantities for the fish subset only. In each case we see that 1st and 2nd position frequencies increase approximately linearly with FFD frequency, but with slopes less than 1. If selection dominated mutation, the base frequencies would be fixed at the values that optimize the amino acid sequence of the protein. Therefore these graphs would be horizontal lines, independent of the FFD frequency. If mutation dominated selection, the base frequencies at 1st and 2nd positions would be equal to FFD frequencies. Hence, the graphs would have a slope of 1. The actual data has a slope between 0 and 1, indicating that both mutation and selection are relevant. The fact that the slopes for 2$^{nd}$ position are less than those for 1$^{st}$ position shows that selection at 2$^{nd}$ position is stronger than at 1$^{st}$ position.

We will consider the simplest possible model that explains the trends in these data and gives a quantitative measure of the strength of selection in different data sets. The model is similar to that used by Sueoka (1998), except that we consider the four bases separately rather than just G+C, and we use the frequency at FFD sites as the independent variable rather than all the 3$^{rd}$ position sites. Let $f_{ik}^{(1)}$, $f_{ik}^{(2)}$ and $f_{ik}^{(4)}$ be the frequencies of base $k$ in species $i$ at 1$^{st}$ position, 2$^{nd}$ position and FFD sites, respectively. Suppose that there is a fraction $\varepsilon_1$ of 1$^{st}$-position sites where selection is negligible and the base is free to vary in the same way as at FFD sites, and a fraction 1-$\varepsilon_1$ where selection is very strong and the base is not able to vary at all. Let $\phi_k^{(1)}$ be the frequency of base $k$ at the strongly selected sites. The frequency of the bases in each species at 1$^{st}$ position should therefore be:

$$f_{ik}^{(1)} = (1-\varepsilon_1)\phi_k^{(1)} + \varepsilon_1 f_{ik}^{(4)}. \tag{1}$$

Similarly, if the fraction of variable sites at second position is $\varepsilon_2$ and the frequencies of the bases in the strongly selected sites are $\phi_k^{(2)}$, then the 2$^{nd}$ position frequencies in each species will be:

$$f_{ik}^{(2)} = (1-\varepsilon_2)\phi_k^{(2)} + \varepsilon_2 f_{ik}^{(4)}. \tag{2}$$

This predicts that the graphs in Figures 2 and 3 should be straight lines. The lines shown in the figures are least-squares fits to the model. The four lines for 1st position are fitted simultaneously (see Appendix A for more details). There is a single $\varepsilon_1$ parameter and three independent $\phi_k^{(1)}$ parameters (because of the constraint that the four $\phi_k^{(1)}$ values must sum to 1). Thus there are four parameters for all four 1$^{st}$-position graphs, whereas simple linear regression of the four graphs independently would require eight parameters. The four graphs for second position are also fitted simultaneously in the same way.

The model predicts the trends well but there is considerable scatter in the data points. In fact, the metazoan data set is very diverse, and it may be unreasonable to assume that a single set of model parameters apply to the whole set. We therefore analyzed the fish and mammal data sets separately. These are the two largest available monophyletic groups of closely-related species. Comparison of the fish data (Fig. 3) with the metazoan data (Fig. 2) shows that there is much less scatter in the smaller data set. The mammal data set is similar to the fish in this respect. In section 7 we will carry out a careful analysis of the scatter in these data points, but firstly we wish to focus on the trends revealed by fitting the model.

Table 1 shows the optimal values of the parameters for the 1st and 2nd positions for each data set. There is a definite difference between the optimal frequencies at the 1st and 2nd positions. The four frequencies at the 1st position are roughly equal in each of the data sets, whereas at the 2nd position there is a high frequency of U, a moderate frequency of C, and low frequencies of A and G. The optimal frequencies are controlled by selection, not mutation. This indicates that selection prefers the use of amino acids whose codons have U or C at the 2nd position, *i.e.* those in the first two columns of the genetic code diagram (see Figure 4).

From Table 1, it is apparent that the $\varepsilon$ parameter varies between different data sets, but for all the data sets, it is higher for the 1st position than the 2nd. The simple interpretation of this is that there is a larger fraction of variable sites at 1st position than 2nd. However, we would like to state the result in a more general way: 1st position sites are more responsive to directional mutation pressure than 2nd position sites. Our interpretation of this is that pairs of amino acids related by 1st position changes tend to be more similar in physical properties than those related by 2nd position changes. Substitutions between pairs of similar amino acids are more likely because selection acts less strongly against them. This allows base frequencies at 1st position to respond more easily to the variation in the frequencies prescribed by the mutation process.

The model is clearly too simple in that it assumes that selection at a site is either negligible or very strong, whereas in reality there will be a spectrum of sites with different selective strengths acting. If a site is under very strong selection, it is not subject to the influence of mutation pressure. If a site is under negligible selection, the base frequencies at that site are free to vary by random drift. If the effective population size, $N_e$, is large, such sites will be polymorphic. If $N_e$ is small, there will be few polymorphic sites, and most of the non-selected sites will be randomly fixed for one base or another. Occasionally the base present at a site will flip due to the fixation of a neutral mutation. The frequencies of bases at non-selected sites will be equal to the equilibrium frequencies of the mutation process, both for polymorphic and randomly fixed sites.

However, the situation is more complex if selection and mutation are both important. Suppose there is a preferred base at a site, such that sequences with this base have fitness $1+s$ times greater than sequences with the other three bases. If $s$ is of comparable magnitude to the per-base mutation rate, then the expected frequencies of bases at this site will depend on many parameters. At fixed sites, the frequencies will depend on the fixation rate of advantageous and deleterious mutations, which is a function of $N_e$, $s$, and the mutation rates. At the polymorphic sites, when $N_e$ is large, the base frequencies will depend on mutation-selection balance. For moderately-selected sites, the full details of the mutation rate matrix become important. For example, the HKY model (Hasegawa *et al.* 1985), used in molecular phylogenetics, is defined so that the rates of transitions from base *i* to base *j* are $r_{ij} = \alpha \pi_j$, whilst the rates of transversions are $r_{ij} = \beta \pi_j$. It is often found that transitions happen faster than transversions ($\alpha > \beta$). For non-selected sites, only the equilibrium frequencies, $\pi_j$ are relevant for determining the observed base frequencies, whereas for moderately selected sites, the values of $\alpha$ and $\beta$ are also important because base frequencies will depend on the relative sizes of mutation rates and selection coefficients.

A full analysis of moderately selected sites would require information from multiple sequences within one species. This would allow identification of polymorphic sites and comparison of sequence divergence within and between species. This can be done for a few well-studied groups of species, such as humans and apes (Hasegawa *et al.* 1999) and *Drosophila* (Dean & Ballard, 2005). In our case we have information on one typical sequence from a large number of species, and we cannot say anything about polymorphisms within species. Although simple, the model proposed above is an effective way to assess the relative importance of mutation and selection on different types of site. The key point is that changes at 1st position sites seem to be under weaker selection that those at 2nd position sites. This has also been shown by Knight *et al.* (2001b). In our previous work on phylogenetics using mitochondrial genes (Gibson *et al.* 2005), we showed that the rate of substitution at 1st position sites is considerably larger than at 2nd position. This is also consistent with there being a weaker selection at 1st position or a larger fraction of possible neutral substitutions at 1st position. A similar pattern is seen in bacterial genomes (Muto & Osawa, 1987; Kimura 1983) when G+C content is considered as the only variable.

## 4. Quantifying amino acid similarities

In this section we develop a measure of similarity between amino acids that will enable us to confirm our explanation for the difference between 1st and 2nd position changes, and allow us to interpret several more detailed observations on the mitochondrial sequence data. We have previously listed a set of 8 physical properties of amino acids that are thought to influence protein structure and folding - see sections 2.4-2.6 of Higgs & Attwood (2005). These properties are volume (Creighton, 1993), bulkiness, polarity and isoelectric point (Zimmerman *et al.* 1968), two different measures of hydrophobicity (Kyte & Doolittle, 1982; Engelman *et al.* 1986), surface area accessible to water (Miller *et al.* 1987), and fraction of accessible area lost when a protein folds (Rose *et al.* 1985). For numerical values of these properties, see Table 2.2 of Higgs & Attwood (2005).

Let $X_{ij}$ be the value of the $j^{th}$ physical property for amino acid $i$. Let $\mu_j$ and $\sigma_j$ be the mean and standard deviation of the $j^{th}$ property, respectively. We define normalized quantities $z_{ij}$ so that each property is on a comparable scale, independent of the original units:

$$z_{ij} = \frac{X_{ij} - \mu_j}{\sigma_j}. \tag{3}$$

Hence, we define the distance between any two amino acids $i$ and $j$ as the euclidean distance between the points in the eight-dimensional space of the $z$ coordinates:

$$d_{ij} = \left( \sum_{k=1}^{8} (z_{ik} - z_{jk})^2 \right)^{1/2}. \tag{4}$$

The distance matrix calculated in this way is given in Appendix B of this paper. The most similar pair of amino acids is I and L, with $d_{IL} = 0.368$, and the most distant pair is R and G, with $d_{RG} = 7.415$.

A convenient way to visualize the distances between amino acids is to use principal component analysis to project the eight dimensional space onto the first two principal components (further details in Higgs & Attwood, 2005). The result is shown in Figure 5. The first component is strongly correlated with hydrophobicity (hydrophobic amino acids are on the left and hydrophilic on the right). The second component is strongly correlated with volume (the largest amino acids are at the top and the smallest at the bottom). The principal component plot demonstrates many of the similarities that we might expect intuitively: the basic amino acids R and K are close, the acids D and E are close, the amines N and Q are close, and there is a tight cluster of hydrophobic amino acids F, L, I, M and V.

The symbols in Figure 5 have been chosen in order to illustrate the relationship between the genetic code structure and the physical properties of the amino acids. The amino acids in column 1 of the code form the tight hydrophobic cluster F, L, I, M and V. Those in column 2 (S, P, T and A) also form a closely related cluster. The majority of the column-3 amino acids (H, Q, N, D and E) fall fairly close to one another in property space, although Y and K are somewhat removed from this group. In contrast, it is striking that the column-4 amino acids, C, W, R and G, are very far from one another. There also seems to be no obvious similarity between amino acids in the same row.

## 5. Proximity and Responsiveness

We showed in section 3 that $1^{st}$ position sites are more responsive (*i.e.* the slope parameter $\varepsilon$ is higher) than $2^{nd}$ position sites, and we predicted that this is because amino acids related by $1^{st}$ position changes are more similar than those related by $2^{nd}$ position changes. This can now be confirmed. The simplest measure of similarity that can be obtained from the distance matrix is the inverse distance. We will define the 'proximity' of amino acids $i$ and $j$ as $1/d_{ij}$. In Table 2, we show the mean proximity of all amino acid changes caused by $1^{st}$ position substitutions and the mean of those caused by $2^{nd}$ position substitutions. In these averages, all non-synonymous single-base substitutions were weighted equally, and substitutions involving stop codons were excluded. As expected, the $1^{st}$ position proximity value is significantly larger.

Although on average, amino acids related by $1^{st}$ position changes are more similar than for $2^{nd}$ position, Fig. 5 reveals that this is principally due to the high similarity of the amino acids in the $1^{st}$ and $2^{nd}$ columns of the genetic code. Column 3 amino acids are less similar, and column 4 amino acids are not similar at all. Table 2 shows the mean proximities for substitutions in each of the four columns. Each $1^{st}$ position non-synonymous substitution in a given column of the code is weighted equally, and changes involving stop codons are excluded from the average. This reveals a clear trend that col. 1 > col. 2 > col. 3 > col. 4. From this we predict that, in the mitochondrial sequence data, the responsiveness of $1^{st}$ position changes to directional mutation pressure should depend on the base present at the $2^{nd}$ codon position. Note that, in column 1, synonymous $1^{st}$ position changes are possible between leucine codons. These were excluded from the average because $1/d_{ij}$ would be infinite. This means that the predicted high proximity and responsiveness of column 1 are due to the properties of the full set of amino acids in this column, and not simply due to the possibility of synonymous changes at 1$^{st}$ position in this column.

We measured the frequencies of 1$^{st}$ position bases $f_{ik}^{(1)}$ for each species, as in section 3, but we only counted codons with a specified 2$^{nd}$ base. Using the same model as in section 3, we fitted the 1$^{st}$ position data to the variation in FFD site frequencies separately for each of the four columns on the genetic code. Table 3 shows these results for the fish and mammal data sets. (The complete Metazoa set was not analyzed this way because there are several variant genetic codes within these species, whereas both fish and mammals use the vertebrate code shown in Figure 4.) Figure 6 shows the fit of U1 against U4 in each of the four columns for the fish data. This emphasizes that the four columns are clearly not equivalent.

For the fish, the responsiveness $\varepsilon$ in Table 3 shows a clear decreasing trend from 1$^{st}$ to 4$^{th}$ column, as expected from the proximity values in Table 2. For the mammals, the 1$^{st}$ column is clearly the largest and the 4$^{th}$ the smallest, but the 2$^{nd}$ and 3$^{rd}$ are equal, which is broadly consistent with expectations. Both the column 4 slopes appear slightly negative. This is probably because 1$^{st}$ and 2$^{nd}$ position changes are not isolated from one another. In column 4, the amino acids are very different from one another; therefore 1$^{st}$ position changes are rare. The relative frequency of codons in column 4 could therefore be influenced by the relatively rare changes at second position, *i.e.* changes into and out of column 4. This could lead to slight trends in the opposite direction to those expected. In the other three columns, changes at the first position are easier than those at second. Therefore the trends follow the direction expected from considering first position changes only. Note that the mean proximity value for column 4 is lower than that for 2$^{nd}$ position, whereas for the other three columns, it is higher (see Table 2).

## 6. Variation of Amino Acid Frequencies

As the frequencies of bases at 1$^{st}$ and 2$^{nd}$ positions vary, there is a corresponding variation in amino acid frequencies in the proteins coded by these genes. We therefore wish to examine the response of the individual amino acid frequencies to directional mutation pressure. We also saw in the previous section that 1$^{st}$ and 2$^{nd}$ position base changes cannot strictly be treated in isolation. Amino acid frequencies will reflect the simultaneous changes in frequencies of bases at all positions.

Figure 7 shows examples of the way in which three amino acids respond to changes in U4. This figure applies to the fish data set, and (as with all the results in this paper) includes only the sequences of the genes on the plus strand of the genomes. It can be seen that serine (counting UCN codons only) shows a systematic increase in frequency with U4 and that threonine shows a systematic decrease. Although the frequency of alanine fluctuates considerably between species, there is no systematic trend with U4.

The trends of the data points on plots like those in Figure 7 can be summarized by measuring the slope of the linear regression line of each amino acid against each of the N4 base frequencies. These slopes are shown in Figure 8 for the fish and mammals data sets. The amino acids have been listed in the order that they appear in the genetic code, moving from column 1 through to column 4. Serine UCN and AGY are treated as two separate amino acids because these blocks of codons are not accessible from one another by a mutation at a single position and they respond in different ways to mutation pressure. Leucine also has six codons, but these are all accessible by a single mutation at first or third position. We have therefore treated all the leucine codons as a single group.

The fish and the mammal datasets are independent of one another but they tend to respond in similar ways to mutation pressure. The slopes in the two sets are correlated: the Pearson correlation coefficient is R = 0.76, p < $10^{-16}$. Nevertheless, it can be seen in Figure 8 that there are a few cases where the response of an amino acid is in opposite directions in the two data sets.

The amino acids in the first two columns of the genetic code (from F to A) tend to have large slopes (either positive or negative), whilst those in the third and fourth columns (from Y to G) tend to have slopes close to zero. To quantify this, we will define the responsiveness of an amino acid to be the root mean square value of the 8 slopes for that amino acid (i.e. the slopes against each of the 4 bases for both the fish and mammal data sets). Responsiveness values for each amino acid are quoted in Table 4. Responsiveness measures the ease with which the amino acid frequency responds to directional mutation pressure acting on base frequencies. The most responsive amino acids are I, V and L, and the least responsive are W, R and D.

We will now use the distance matrix $d_{ij}$ to predict the responsiveness of the different amino acids. Our hypothesis is that an amino acid frequency should respond more to mutation pressure if there are other amino acids with similar properties that are accessible via single mutations. We define the proximity value for an amino acid as the mean of $1/d_{ij}$ for all codons accessible by single non-synonymous substitutions from the codon set for that amino acid. Substitutions at all three codon positions are included in this average, and we also include stop codons here. These are treated as being at infinite distance ($1/d = 0$). As an example, consider threonine (T):

$$\text{proximity(T)} = \frac{1}{24}\left(\frac{2}{d_{TI}} + \frac{2}{d_{TM}} + \frac{6}{d_{TS}} + \frac{4}{d_{TP}} + \frac{4}{d_{TA}} + \frac{2}{d_{TN}} + \frac{2}{d_{TK}} + 2\times 0\right) \quad (5)$$

Note that there are 2 codons for I and M accessible from the T codon block, 6 for S, *etc.* The $2\times 0$ represents the stop codons. We define the proximity separately for the two serine codon blocks, as we did for the responsiveness.

Figure 9 shows that there is a strong correlation between responsiveness and proximity: rank correlation coefficient R = 0.85, p < $10^{-6}$ (see Table 5). This confirms the hypothesis that the physical properties and the genetic code structure influence the responsiveness of the amino acids, and shows that we can successfully predict responsiveness from the physical property distance matrix.

Another related quantity is the mutability of amino acids measured from amino acid substitution rate matrices. A variety of substitution rate matrices have been defined, stemming from the initial PAM matrices (Dayhoff *et al.* 1978; Jones *et al.* 1992). The one most relevant to this study is the mtREV matrix of Adachi & Hasegawa (1996), since it is derived from vertebrate mitochondrial protein sequences. We define the mutability of an amino acid as the net rate of change from that amino acid to all others, normalized so that the average rate is 1. Table 3 of Adachi & Hasegawa (1996) gives the matrix of probabilities $P_{ij}$ that amino acid $i$ is replaced by amino acid $j$ in a time such that the average probability of substitution is 1/100. The probability that an amino acid does not change in this time is $P_{ii}$. The mutability is $m_i = 100(1-P_{ii})$. Mutability values are given in Table 4. The mtREV matrix considers amino acids, not codons; therefore, there is only one figure for serine, and we have entered the same figure for both blocks of serine codons in the table. We find that mutability is significantly correlated with both responsiveness and proximity (see Table 5). Mutability and responsiveness are thus influenced by the physical properties of the amino acids in similar ways.

## 7. Data Point Scatter

In Figure 2 and 3 there is considerable scatter of data points about the straight lines. A certain amount of this is due to random sampling effects in finite length sequences. However large deviations may be indicative of effects not described by the theory, or of heterogeneity in the data. Here we will analyze the distribution of deviations between data points and theory. This reveals important information about the nature of selection on the different types of site.

For any species $i$, let $l_{ik}$, $m_{ik}$ and $n_{ik}$ be the number of observed occurrences of base $k$ at FFD, 1$^{st}$ and 2$^{nd}$ position sites, respectively. Let $L = \sum_k l_{ik}$, $M = \sum_k m_{ik}$ and $N = \sum_k n_{ik}$. We suppose that the FFD sites are generated by a mutational process with equilibrium base frequencies $\pi_{ik}$. The estimated values of these frequencies are $\hat{\pi}_{ik} = l_{ik} / L_i$. According to the model in section 3, the expected frequencies of the bases at the 1$^{st}$ and 2$^{nd}$ positions, are:

$$f_{ik}^{(1)} = (1 - \varepsilon_1)\phi_k^{(1)} + \varepsilon_1 \hat{\pi}_{ik}, \tag{6a}$$

and

$$f_{ik}^{(2)} = (1 - \varepsilon_2)\phi_k^{(2)} + \varepsilon_2 \hat{\pi}_{ik}. \tag{6b}$$

The expected numbers of occurrences of bases at these positions are $m_{ik}^{\exp} = M_i f_{ik}^{(1)}$, and $n_{ik}^{\exp} = N_i f_{ik}^{(2)}$.

The deviations between the observed and expected number of occurrences of the bases at the two positions are:

$$X_i^{(1)} = \sum_k \frac{(m_{ik}^{\exp} - m_{ik})^2}{m_{ik}^{\exp}}, \tag{7a}$$

and

$$X_i^{(2)} = \sum_k \frac{(n_{ik}^{\exp} - n_{ik})^2}{n_{ik}^{\exp}}. \tag{7b}$$

To a good approximation, if the model is valid, the distribution of these deviations should be a chi-squared distribution with three degrees of freedom (because there are four bases and one constraint on the sum of the base frequencies). This is only an approximation for two reasons. Firstly, the $\varepsilon$ and $\phi$ parameters have been estimated from the data. This will make the deviations slightly smaller than if the true parameters were known without looking at the data.

However, each individual species only contributes a small amount to the fitting of the full data set, therefore this effect should be small. Secondly, the $\hat{\pi}_{ik}$ are only estimates of the true frequencies $\pi_{ik}$. The 1st and 2nd position frequencies in equations 6a and 6b should depend on the true frequencies not the estimated ones. This will make the deviations slightly larger than if the true $\pi_{ik}$ were known. We will now show that for simulated data, the distribution is very close to the chi-squared distribution, so neither of the effects discussed here is important. However, the distribution for the real data is significantly different, which indicates that the model does not fully explain the real data.

We generated simulated data in the following way, so to be as close as possible to the real data but to follow the theory exactly. The $\varepsilon$ and $\phi$ parameters were estimated for the real data as in section 3. For each species, the $\hat{\pi}_{ik}$ were estimated as above. These were then used as the true frequencies for the simulated data. Simulated values of $l_{ik}$ were obtained by selecting $L_i$ bases with frequencies $\hat{\pi}_{ik}$. Simulated values of $m_{ik}$ and $n_{ik}$ were obtained by selecting $M_i$ and $N_i$ random bases with the frequencies in equations 6a and 6b. The simulated data was then treated exactly as the real data. New values of the $\varepsilon$ and $\phi$ parameters were obtained by fitting the model to the simulated data. New estimated frequencies $\hat{\pi}_{ik}$ were obtained for the simulated data, and the deviations were calculated as above.

Figure 10 compares the real and the random data for the metazoa and fish data sets. Each graph shows U1 and U2 as a function of U4. For the metazoa, there is noticeably less scatter in the simulated data than the real data for both 1st and 2nd positions. For the fish, there is slightly less scatter in the simulated data for 1st position, but somewhat more scatter for the 2nd position. These things are also visible in Figure 11, where we show the distributions of the deviations. The graphs show the probability F(X) that the measured deviation is greater than or equal to X. Each graph compares F(X) for real and simulated data and the curve from the chi-squared distribution with three degrees of freedom. In each case the simulated data curve is close to the chi-squared curve. For the metazoa, the real data curve is much to the right of the chi-squared curve for both 1st and 2nd positions. This indicates a significantly higher probability of large deviations. This is most likely due to heterogeneity in the broad metazoa dataset. Different groups of species may have different base frequencies in the optimal sequences (*i.e.* the $\phi$ parameters may be different for different groups). Species will also differ in the relative rate of mutation to selection strength. So it may be a poor approximation to insist that all species fall on the same line.

For the fish, the real data curve for position 1 is only slightly to the right of the chi-squared curve, indicating that the model is a much better fit for the narrower data set than for the complete set of metazoa. At position 2, the real data curve is to the left of the chi-squared curve, *i.e.* deviations are smaller than expected from random sampling. We also calculated the distributions for the mammal data, and the graphs are similar to those for the fish.

On reflection, we might have predicted the smaller than expected deviations at position 2. If we take the model in section 3 literally, there is a certain fraction of sites that are fixed, and these should not be subject to random sampling. Only the variable sites will be subject to random sampling, and the fluctuations at 2nd position should therefore be smaller than those at 1st position because the fraction of variable sites is smaller.

The number of variable sites at 2nd position should be $N_i \varepsilon_2$, and these sites should be distributed with frequencies $\pi_{ik}$. The expected number of occurrences of base $k$ at variable sites is therefore

$$n_{ik}^{\exp} = N_i \varepsilon_2 \hat{\pi}_{ik}.$$  (8)

If we subtract the number of occurrences of base $k$ at fixed sites from the total number of occurrences of base $k$, then we obtain the observed number of occurrences of base $k$ at variable sites:

$$n_{ik}^{obs} = n_{ik} - N_i (1 - \varepsilon_2) \phi_k^{(2)}.$$  (9)

In principle, we could construct a chi-squared deviation from the observed and expected numbers in equations 8 and 9. However, this proved to be impossible, because for some species, the observed number calculated from equation 9 turned out to be negative. We conclude from this that the fixed sites cannot literally be fixed in all species. Some degree of fluctuation in base frequencies at the 'fixed' sites must still be occurring; otherwise it would be impossible to obtain negative numbers from equation 9. In summary, the deviations are smaller than we would expect if all position 2 sites were chosen by random sampling, but less than we would expect if the fixed sites were truly fixed and only the variable sites were chosen by random sampling.

As we discussed in section 3, it is an over-simplification to divide sites into completely neutral and completely invariant sites, and it is likely that there is a spectrum of sites with different selection strengths $s$ acting. We could predict the probability of fixation of deleterious bases at a given site as a function of $s$ and the effective population size, $N_e$. This would give some prediction of the range of the scatter to expect. One prediction is that there should be more scatter in species with smaller $N_e$. However, we do not have estimates of $N_e$ for the vast majority of species in our data set.

Although this analysis of the deviations reveals the limitation of the simple model with fixed and variable sites, it qualitatively confirms our interpretation of the data. The key point is that the $2^{nd}$ position sites are more constrained by selection than the 1st position sites; therefore the deviations are smaller at $2^{nd}$ position than $1^{st}$, and the $2^{nd}$ position sites are less responsive to mutation pressure than the $1^{st}$.

## 8. Variation between strands and along the genome

In the majority of this paper, we considered variation of properties between genomes. In this section, we wish to consider variation between the strands of a given genome and variation among the genes on a given strand. To understand this, it is necessary to consider the mechanism of DNA replication more carefully.

In mitochondria, the asymmetry of the mutation rates causing mutation pressure away from equal base frequencies is linked with the asymmetry of replication of the DNA strands. According to the usual understanding of mitochondrial genome replication (Bogenhagen & Clayton, 2003), synthesis of the new H strand begins at an origin site, $O_H$, and proceeds in one direction. The original H strand remains single stranded until another origin site, $O_L$, for the L strand is reached that is more than half way round the genome. At this point, synthesis of the new L strand begins from $O_L$ and proceeds in the other direction. Reyes *et al.* (1998) discuss the deamination reactions from C to U and from A to hypoxanthine that can occur on single stranded regions of DNA. As the H strand is single stranded for longer, there should be a net decrease of C and A, and corresponding increase of U and G on the H strand, whilst on the L strand there should be an increase in C and A and decrease in U and G. The analysis here included the genes for which the H strand is the template (12 out of 13 genes in vertebrates); hence the base frequencies in the sense strands for these genes have the biases predicted for the L strand. This explains why C4 > U4 and A4 > G4 for the majority of species, as can be seen from Figures 1. There are, nevertheless, a considerable number of species that do not follow these rules. The fact that the base frequencies change a lot between species indicates that the rates of the chemical reactions causing the mutations must be different in different species. Further experimental information on these rates would clearly be of use in interpreting the patterns we have seen from the sequence analysis.

It should also be noted that the H and L designation for the strands is tied into the replication mechanism, and we can only use this notation for vertebrates, where a considerable amount is known about the replication process, and where $O_H$ and $O_L$ sites have been identified. In invertebrates, these sites are not often identified in genome annotations, and it is possible that replication mechanisms may differ. There has also been controversy recently regarding the replication mechanism in vertebrates (Bowmaker *et al.* 2003; Bogenhagen & Clayton, 2003), and the model of unidirectional strand replication seems less certain than it once did. Another complication is that there are frequent gene inversions seen in invertebrate genomes, and this means that there is no straightforward relationship between one strand in a vertebrate genome and one strand in a genome from another animal phylum. For this reason, we have preferred to designate the strands simply as plus and minus in the present paper, where the plus strand is the sense strand for the majority of genes. The occurrence of inversions might influence base frequencies in individual genes in an interesting way, because a gene that changes strands would find itself out of equilibrium with the mutational process on the new strand. This is an effect that we could potentially consider in future. However, we note that there are substantial base frequency variations among both the fish set and the mammal set where all the protein coding genes remain on the same strand and gene order is unchanged. These variations must really be due to changes in the mutation rates among species and cannot be influenced by genes switching strands.

According to the asymmetrical, unidirectional mechanism of genome replication, genes encoded on the H strand spend a variable amount of time in a singe stranded state, depending on their position on the genome relative to the replication origin. A quantity $D_{ssH}$ has been proposed as a measure of the amount of time a gene is single stranded (Reyes *et al.* 1998; Faith & Pollock, 2003). If mutations occurring in the single stranded state are responsible for the asymmetry of the mutation process, we would expect base frequencies to vary along the genome according to the amount of time spent single stranded. In fact, base frequencies in individual genes are found to correlate with $D_{ssH}$ (Reyes *et al.* 1998; Faith & Pollock, 2003; Gibson *et al.*

2004), which is consistent with the asymmetrical, unidirectional genome replication mechanism. In a phylogenetic context, maximum likelihood models have also been developed that incorporate strand asymmetry (Bielawski & Gold, 2002; Krishnan *et al.* 2004). Raina *et al.* (2005) have also shown that strand asymmetry at $1^{st}$ and $2^{nd}$ positions is related to that at FFD sites in a maximum likelihood phylogenetic study of primates.

We now wish to compare the size of the variation in base frequency along a genome with that between genomes. The number of occurrences of each base at FFD sites and at $1^{st}$ and $2^{nd}$ positions in each gene in each of the 109 mammalian genomes was counted. These counts were summed over genomes and then divided by the total number of counts at each type of site, resulting in average frequencies in each gene for all the mammals. These frequencies are shown in Figure 12. The genes are ranked in order of increasing $D_{ssH}$ (following Reyes *et al.* 1998). Since the protein coding genes are in the same order on all the mammalian genomes, this ranking is the same for all the genomes included in the average. For the FFD sites in the plus-strand genes (open circles), linear regression lines are shown. All these lines have slopes significantly different from zero (p< 0.001 for U and C, p < 0.005 for G, p < 0.01 for A). This confirms that there is a significant variation in the mutational process along the genome, as has already been seen (Reyes *et al.* 1998; Faith & Pollock, 2003; Gibson *et al.* 2005). Following the arguments given in the main part of this paper, we would expect this to cause a corresponding variation in frequencies at $1^{st}$ and $2^{nd}$ positions, *i.e.* the $1^{st}$ and $2^{nd}$ position points in Figure 12 should follow a trend in the same direction as the corresponding base at the FFD sites, but with smaller slopes. In fact none of the slopes of the linear regression lines for $1^{st}$ and $2^{nd}$ position points is significantly different from zero. This is at least consistent with the argument that selection limits the variation that can occur at $1^{st}$ and $2^{nd}$ position, but it is not possible to test whether the $1^{st}$ position slope is greater than the $2^{nd}$ position slope. The reason for the lack of significance of the test on the regression slopes is that there is a large scatter of the points between genes caused by the variation in the amino acids required in each gene. The majority of the $1^{st}$ and $2^{nd}$ position base frequency variation between genes is thus due to the specific amino acid sequences of the proteins and not due to the underlying mutational trend. Although the mutational variation along the genome is significant (as seen from the FFD sites), it is a smaller effect than the variation between genomes. Table 6 compares the minimum and maximum FFD genome frequencies with the minimum and maximum FFD gene frequencies, and shows that the within-genome range is narrower than the across-genome range. The across genome range would be wider still if species outside the mammals were included.

Although the variation between the genes on the plus strand is relatively small, the variation between plus and minus strand genes is larger. This can also be seen in Figure 12 by comparing open points and filled points. To see this effect in more detail, we calculated the 1st position, 2nd position and FFD site base frequencies on both strands for each genome. The plus strand frequencies are averages of 12 genes, and the minus strand frequencies are for ND6 only. We then used the model in Section 3 to fit the data for each genome separately. For each base and for each of 1st and 2nd positions, there are just two points corresponding to the two strands, rather than one point for each species. The formulae for fitting the model are the same as in Appendix A, expect that the sums over species are replaced by sums over the two strands in one species. The key parameters to be estimated are the slope parameters $\varepsilon_1$ and $\varepsilon_2$, which determine the degree to which 1st and 2nd position base frequencies are able to diverge between the strands as a result of the divergent mutational process acting on the two strands.

We expect that the divergence between the strands should be governed by the same selective processes that influence variation between species. Therefore, we expect that both slopes should be positive (i.e. the trend between strands at 1st and 2nd positions should be in the same direction as the trend at the FFD sites), and that $\varepsilon_1 > \varepsilon_2$ (i.e. selection is stronger at 2nd position). For the 109 mammals analyzed, these conditions were true for every species. For the 172 fish analyzed, the conditions were true for all but one species. The final fish species, *Albula glossodonta*, differs significantly from expectations, in that both $\varepsilon_1$ and $\varepsilon_2$ are negative, *i.e.* the frequencies at 1st and 2nd positions vary in opposition to those at the FFD sites. This does not make sense if the base frequencies are in equilibrium under mutation and selection. The most likely explanation seems to be that there has been a recent sudden change in the mutation processes in this species, FFD sites have changed rapidly in response to this, but 1st and 2nd position sites are still out of equilibrium and reflect the mutation process at some point in the past. However, the main point is that, with the exception of this one species, the between strand trends are consistent with our expectations from the cross-genome analysis.

**9. Discussion**

Variation in frequencies of bases and amino acids among sequences is an important source of bias in phylogenetic studies (Foster & Hickey, 1999; Schmitz *et al.* 2002). Gibson *et al.* (2005) considered the proteins in mammalian mitochondrial genomes in detail, and showed that the bases that are most variable in frequency are C and T. For this reason, they proposed a three-state model in which C and T were treated as a single state. This model was shown to remove a number of important discrepancies in the mammalian tree. This study also showed that the rate of substitutions at first position sites is considerably larger than that at second positions. This is another manifestation of the fact that substitutions at the first position change the amino acid properties less than those at the second position. Generally, in phylogenetic studies, it would seem beneficial to treat first and second position sites as separate sets with independent substitution rate matrices. However, even if this is done, single site models cannot capture detailed effects such as the greater frequency of first position substitutions within the first and second column of the genetic code than within the third and fourth columns. More complex effects can be treated better by codon-based models, such as those of Goldman & Yang (1994 and Halpern & Bruno (1998). The first of these incorporates an amino acid distance matrix given by Grantham (1974), which is similar in spirit to the one we used here.

An important assumption in fitting the data is that the base frequencies at FFD sites are determined by the equilibrium frequencies of the mutational process. We suppose that the equilibrium frequencies gradually vary along lineages and that the observed frequencies at FFD sites are tracking this variation. If the mutation rate were low, then the observed frequencies would lag behind the equilibrium frequencies. This may be one factor contributing to the excess scatter observed in the data. Nevertheless, these results show that the mutation rate is large enough to drive both $1^{st}$ and $2^{nd}$ position base frequencies and amino acid frequencies away from their optimal values. Since mutation rates are large, it is likely that the observed frequencies do not differ much from the equilibrium ones in most species. However, the example of *Albula glossodonta* in section 8 seems to be a case where base frequencies are out of equilibrium. It would be of interest to look specifically for non-equilibrium effects in base frequencies, possibly by close analysis of small groups of species where a phylogenetic tree is available, and where base frequency changes can be localized on particular branches of the tree, as has been done with primates by Raina *et al.* (2005).

Another important assumption was that selective effects acting on FFD sites were negligible. Cases of codon bias have been detected in many organisms (*e.g.* Kanaya *et al.* 1999; Coghlan & Wolfe, 2000; Duret 2000), and this is often explained in terms of selection for using the codons that match the most common tRNAs. In animal mitochondrial genomes, however, there is only one tRNA for each codon family. With a very small number of exceptions, four codon families always have a tRNA with a U in the wobble position, two codon families ending in U and C always have a G in the tRNA wobble position, and two-codon families ending in A and G always have a U in the tRNA wobble position. These features of the tRNAs are constrained by the fact that one tRNA has to pair with all codons in the codon family. Since the tRNA anticodons do not vary between species they cannot cause varying selective pressure in different organisms and different amino acid groups, in the way that they do in bacteria and some larger organisms. Separately from any tRNA-related effects, it is also possible that selective effects at the DNA level may influence the choice of bases at the FFD sites (Antezana & Kreitman, 1999). A further very specific case where weak selection at FFD sites might arise is in leucine codons: CUN and UUY. It is possible for a codon coding for leucine to synonymously mutate into a form where the third codon positions are under selection and then mutate back, allowing selection to act on nucleotides at the FFD site. A detailed analysis of codon usage in the mitochondrial data would certainly be of interest, but it is beyond the scope of the current paper. The picture emerging from our results is that the mutation rate is large, and causes significant variation even at nonsynonymous sites; therefore it seems reasonable to neglect selection at synonymous sites for the present analysis.

It has been shown previously that the arrangement of the amino acids within the genetic code table is far from random, and that neighbouring amino acids tend to have similar properties. As a result, the effect of deleterious mutations and errors in translation is reduced with respect to other hypothetical codes in which the positions of the amino acids are reshuffled. It has therefore been argued that the canonical code has been optimized to reduce the severity of these types of errors (Woese, 1965; Alff-Steinberger, 1969; Haig & Hurst, 1991; Freeland *et al.* 2000; Gilis *et al.* 2001). These studies consider a measure of distance between amino acids, and then define an error function that is an average of the distance measure for all possible single base substitutions. The canonical code has a smaller value of the error function than almost all randomized codes. The significance of this result depends on the details of the distance measure and the error function; however, it has been shown several times in different ways, and it appears to be robust.

The most frequently used physical property in the genetic code literature is the polar requirement measure of Woese *et al.* (1966), and the distance between amino acids is then simply the difference in polar requirement (*i.e.* a one-dimensional scale). In this study, we used 8 properties and measured a distance in 8-dimensional space. It seems reasonable that a distance based on a number of properties can better reflect similarities and differences between amino acids than would any one single property. Clearly, our 8 properties are not the only ones that could have been used. We chose these particular properties from the protein folding literature and used them originally as test case in principal component analysis and clustering methods (Higgs & Attwood, 2005). The principal component analysis with these 8 properties brought out similarities between the amino acids in a meaningful way, and we therefore used the same set of properties in the present study. The fact that there is a strong correlation between proximity and responsiveness in Figure 9 shows that this distance measure is a useful one. Woese's polar requirement is not included in the 8 properties, although the properties do include other measures of polarity and hydrophobicity that are correlated with polar requirement. It is likely that the canonical code would also appear optimized with respect to reshuffled codes if our distance measure were used in the error function, but we have not verified this.

Our own recent work on genetic code evolution (Sengupta & Higgs, 2005) is concerned with the origin of variant genetic codes, such as those in mitochondrial genomes, and the mechanism of reassignment of codons to new amino acids. We have proposed four distinct mechanisms for codon reassignment and shown that they can all occur within the same framework. However, the aim of the present paper was not to investigate genetic code evolution, but to explain the response of amino acid frequencies to directional mutation pressure. The layout of the genetic code clearly plays a major part in this. The columns of the code are particularly important since they contain groups of closely related amino acids. The fact that first and third position changes tend to be much more conservative in amino acid properties than second position changes was pointed out in early works on the genetic code (Woese, 1965; Alff-Steinberger, 1969), but was not considered previously in the context of mutation pressure. Our study develops this point to explain why the first position base frequencies respond more easily to mutation pressure than second position changes, and why certain amino acids respond more easily than others.

The distance measure we used is the simplest one obtainable from the 8 pre-selected properties. It weights all of these properties equally and treats them all independently. There is no guarantee that all these properties are equally important to natural selection. It would be possible to assign weights to the different properties in the distance measure, and then optimize the weights so that there was maximal correlation between the proximity and responsiveness of the amino acids in Figure 9. The properties with the largest weights would then be those that are seen as most important by natural selection. Similarly, there is no guarantee that responsiveness should depend on $1/d$ rather than some other decreasing function like $1/d^2$ or $e^{-d}$. We could have tried several other functions to see which gave the best correlation with responsiveness. However, we did not do either of these things, because to do so we would have to assume the result we are trying to prove. We emphasize that the distance measure used here was already developed for a different purpose by Higgs & Attwood (2005) prior to analysis of this data. The fact that the responsiveness of the amino acids can be predicted from this distance measure demonstrates that responsiveness is really dependent on the physical properties of the amino acids.

We did, however, try one important variation on the distance measure. We note that the eight properties are not independent of one another, and there are significant correlations between some pairs of properties. It is for this reason that the principal component analysis reveals significant structure when the data are projected into only two dimensions. Let $\mathbf{S}$ be the matrix of correlation coefficients between the eight properties, and let $\mathbf{S^{-1}}$ be its inverse. The Mahalanobis distance between amino acids $i$ and $j$ is defined as $d_{ij} = \left( (\mathbf{z}_i - \mathbf{z}_j)^T . \mathbf{S}^{-1} . (\mathbf{z}_i - \mathbf{z}_j) \right)^{1/2}$. This is the same as the euclidean distance in equation 4 if there is no correlation between the properties. We calculated the Mahalanobis distances and then used these to determine the proximities. However, this proved not to be useful for the current data. The decrease in mean proximity moving from column1 to column 4 (shown in Table 2) is very clear with the euclidean distance but was found to be less so with the Mahalanobis distance. The rank correlation between the amino acid responsiveness and proximity (shown in Figure 9) is reduced to 0.71 when the proximities are calculated with the Mahalanobis distance, whereas it was 0.85 with the euclidean distance. The Mahalanobis distance is therefore more complicated to calculate and also less useful as a predictor than the euclidean distance. We presume that by down-weighting properties that are correlated with one another, the Mahalanobis distance has also down-weighted the properties that are most important to selection.

In summary, the quantities in Table 4 are derived from different types of information: proximity is dependent on physical properties; responsiveness is measured from the slopes of the amino acid frequencies in the mitochondrial sequence data; mutability is a property of the substitution rate matrix estimated by maximum likelihood. It is therefore gratifying that these quantities give a coherent picture of the differences in behaviour of the amino acids.

## Acknowledgements

This work was supported by the Natural Sciences and Engineering Research Council of Canada and by Canada Research Chairs.

## Appendix A: Least-squares fitting

For a simultaneous least-squares fit of the four data sets for position 1, the quantity to be minimized is

$$S = \sum_{k=1}^{4} \sum_i \left( f_{ik}^{(1)} - (1 - \varepsilon_1) \phi_k^{(1)} - \varepsilon_1 f_{ik}^{(4)} \right)^2 .$$

Let

$$\overline{F_k^{(4)}} = \frac{1}{N_{spec}} \sum_i f_{ik}^{(4)} \; ;$$

$$\overline{F_k^{(1)}} = \frac{1}{N_{spec}} \sum_i f_{ik}^{(1)} \; ;$$

$$\overline{(F_k^{(4)})^2} = \frac{1}{N_{spec}} \sum_i (f_{ik}^{(4)})^2 \; ;$$

$$\overline{F_k^{(4)} F_k^{(1)}} = \frac{1}{N_{spec}} \sum_i f_{ik}^{(4)} f_{ik}^{(1)} \; .$$

By differentiating with respect to the free parameters, it can be shown that $S$ is minimized when

$$\varepsilon_1 = \frac{\sum_k \left( \overline{F_k^{(4)} F_k^{(1)}} - \overline{F_k^{(4)}} . \overline{F_k^{(1)}} \right)}{\sum_k \left( \overline{\left(F_k^{(4)}\right)^2} - \left(\overline{F_k^{(4)}}\right)^2 \right)} \; ;$$

$$\phi_k^{(1)} = \frac{\overline{F_k^{(1)}} - \varepsilon_1 \overline{F_k^{(4)}}}{1 - \varepsilon_1} \; .$$

Note that this solution satisfies the requirement that $\sum_k \phi_k^{(1)} = 1$.

Equivalent formulae apply for the second position.

Appendix B - Table of distances between amino acids derived from 8 physical properties

|   | F | L | I | M | V | S | P | T | A | Y | H | Q | N | K | D | E | C | W | R | G |
|---|---|---|---|---|---|---|---|---|---|---|---|---|---|---|---|---|---|---|---|---|
| F | 0 | 1.160 | 1.189 | 1.031 | 1.855 | 4.544 | 3.675 | 3.166 | 4.11 | 2.062 | 4.093 | 3.889 | 4.402 | 5.888 | 5.758 | 5.171 | 2.904 | 1.982 | 6.269 | 5.958 |
| L | 1.160 | 0 | 0.368 | 1.435 | 0.831 | 4.189 | 3.209 | 2.770 | 3.562 | 2.543 | 4.168 | 3.892 | 4.251 | 5.819 | 5.662 | 5.202 | 2.554 | 2.801 | 6.354 | 5.595 |
| I | 1.189 | 0.368 | 0 | 1.565 | 0.873 | 4.420 | 3.531 | 3.064 | 3.721 | 2.805 | 4.367 | 4.214 | 4.552 | 6.081 | 5.912 | 5.463 | 2.592 | 2.935 | 6.559 | 5.744 |
| M | 1.031 | 1.435 | 1.565 | 0 | 1.879 | 3.618 | 3.014 | 2.404 | 3.263 | 1.853 | 3.513 | 3.243 | 3.624 | 5.456 | 5.152 | 4.657 | 2.217 | 2.440 | 5.905 | 4.995 |
| V | 1.855 | 0.831 | 0.873 | 1.879 | 0 | 3.936 | 3.075 | 2.627 | 3.125 | 3.100 | 4.346 | 4.074 | 4.245 | 6.076 | 5.616 | 5.296 | 2.168 | 3.532 | 6.672 | 5.220 |
| S | 4.544 | 4.189 | 4.420 | 3.618 | 3.936 | 0 | 1.986 | 1.697 | 1.295 | 4.033 | 3.900 | 2.764 | 1.993 | 5.291 | 3.768 | 3.990 | 2.758 | 5.495 | 6.229 | 1.908 |
| P | 3.675 | 3.209 | 3.531 | 3.014 | 3.075 | 1.986 | 0 | 0.869 | 2.276 | 2.909 | 3.357 | 1.882 | 1.759 | 4.375 | 3.864 | 3.688 | 3.047 | 4.292 | 5.425 | 3.787 |
| T | 3.166 | 2.770 | 3.064 | 2.404 | 2.627 | 1.697 | 0.869 | 0 | 1.795 | 2.691 | 3.308 | 2.070 | 1.898 | 4.818 | 3.858 | 3.701 | 2.279 | 4.045 | 5.740 | 3.467 |
| A | 4.110 | 3.562 | 3.721 | 3.263 | 3.125 | 1.295 | 2.276 | 1.795 | 0 | 4.177 | 4.199 | 3.489 | 2.892 | 5.847 | 4.471 | 4.661 | 1.919 | 5.441 | 6.687 | 2.163 |
| Y | 2.062 | 2.543 | 2.805 | 1.853 | 3.100 | 4.033 | 2.909 | 2.691 | 4.177 | 0 | 3.105 | 2.240 | 3.077 | 4.398 | 4.633 | 3.882 | 3.591 | 1.643 | 4.837 | 5.704 |
| H | 4.093 | 4.168 | 4.367 | 3.513 | 4.346 | 3.900 | 3.357 | 3.308 | 4.199 | 3.105 | 0 | 2.912 | 3.164 | 3.027 | 3.692 | 3.134 | 4.094 | 4.033 | 3.352 | 5.115 |
| Q | 3.889 | 3.892 | 4.214 | 3.243 | 4.074 | 2.764 | 1.882 | 2.07 | 3.489 | 2.240 | 2.912 | 0 | 1.090 | 3.595 | 3.225 | 2.743 | 3.868 | 3.835 | 4.462 | 4.496 |
| N | 4.402 | 4.251 | 4.552 | 3.624 | 4.245 | 1.993 | 1.759 | 1.898 | 2.892 | 3.077 | 3.164 | 1.090 | 0 | 4.074 | 2.825 | 2.738 | 3.592 | 4.686 | 4.961 | 3.593 |
| K | 5.888 | 5.819 | 6.081 | 5.456 | 6.076 | 5.291 | 4.375 | 4.818 | 5.847 | 4.398 | 3.027 | 3.595 | 4.074 | 0 | 4.718 | 4.075 | 6.352 | 5.343 | 1.694 | 6.646 |
| D | 5.758 | 5.662 | 5.912 | 5.152 | 5.616 | 3.768 | 3.864 | 3.858 | 4.471 | 4.633 | 3.692 | 3.225 | 2.825 | 4.718 | 0 | 1.109 | 4.849 | 6.010 | 5.541 | 4.827 |
| E | 5.171 | 5.202 | 5.463 | 4.657 | 5.296 | 3.990 | 3.688 | 3.701 | 4.661 | 3.882 | 3.134 | 2.743 | 2.738 | 4.075 | 1.109 | 0 | 4.835 | 5.172 | 4.870 | 5.328 |
| C | 2.904 | 2.554 | 2.592 | 2.217 | 2.168 | 2.758 | 3.047 | 2.279 | 1.919 | 3.591 | 4.094 | 3.868 | 3.592 | 6.352 | 4.849 | 4.835 | 0 | 4.476 | 6.878 | 3.536 |
| W | 1.982 | 2.801 | 2.935 | 2.440 | 3.532 | 5.495 | 4.292 | 4.045 | 5.441 | 1.643 | 4.033 | 3.835 | 4.686 | 5.343 | 6.010 | 5.172 | 4.476 | 0 | 5.516 | 7.080 |
| R | 6.269 | 6.354 | 6.559 | 5.905 | 6.672 | 6.229 | 5.425 | 5.740 | 6.687 | 4.837 | 3.352 | 4.462 | 4.961 | 1.694 | 5.541 | 4.870 | 6.878 | 5.516 | 0 | 7.415 |
| G | 5.958 | 5.595 | 5.744 | 4.995 | 5.220 | 1.908 | 3.787 | 3.467 | 2.163 | 5.704 | 5.115 | 4.496 | 3.593 | 6.646 | 4.827 | 5.328 | 3.536 | 7.080 | 7.415 | 0 |

Table 1. Optimal parameters from fitting the model to the first and second position data.

|  |  | $\varepsilon$ | $\phi_U$ | $\phi_C$ | $\phi_A$ | $\phi_G$ |
|---|---|---|---|---|---|---|
| Metazoa | 1st Position | 0.392 | 24.0 | 21.3 | 22.5 | 32.1 |
| Metazoa | 2nd Position | 0.192 | 46.1 | 25.4 | 13.8 | 14.7 |
| Fish | 1st Position | 0.164 | 20.8 | 25.9 | 24.8 | 28.4 |
| Fish | 2nd Position | 0.039 | 41.1 | 27.7 | 17.7 | 13.5 |
| Mammals | 1st Position | 0.191 | 23.2 | 24.6 | 28.4 | 23.8 |
| Mammals | 2nd Position | 0.066 | 43.5 | 26.7 | 17.6 | 12.2 |

Table 2. Table of mean Proximity for 1st and 2nd positions and for each of the four columns of the genetic code.

| Category | Proximity |
|---|---|
| 1st Position | 0.520 |
| 2nd Position | 0.291 |
| Column 1 | 1.058 |
| Column 2 | 0.669 |
| Column 3 | 0.299 |
| Column 4 | 0.230 |

-----------------------------------------------------------------------

Table 3. Optimal parameters from fitting the model to the first position frequencies for codons in each of the four columns of the genetic code.

| | | $\varepsilon$ | $\phi_U$ | $\phi_C$ | $\phi_A$ | $\phi_G$ |
|---|---|---|---|---|---|---|
| Fish | Column 1 | 0.311 | 23.2 | 34.7 | 26.0 | 16.1 |
| | Column 2 | 0.094 | 17.7 | 19.5 | 28.4 | 34.5 |
| | Column 3 | 0.070 | 15.2 | 29.9 | 28.6 | 26.3 |
| | Column 4 | -0.016 | 29.1 | 15.2 | 11.0 | 44.7 |
| | | | | | | |
| Mammals | Column 1 | 0.396 | 24.9 | 33.0 | 28.6 | 13.6 |
| | Column 2 | 0.060 | 23.1 | 19.6 | 31.5 | 25.9 |
| | Column 3 | 0.060 | 17.5 | 26.3 | 34.1 | 22.1 |
| | Column 4 | -0.001 | 29.1 | 14.9 | 11.5 | 44.5 |

Table 4 - Comparison of amino acid properties. Proximity is calculated from the physical properties of amino acids (Section 4). Responsiveness measures the change in amino acid frequency due to mutation pressure (Section 4). Mutability is calculated from the mtREV substitution rate matrix (Adachi & Hasegawa, 1996).

|        | Proximity | Responsiveness | Mutability |
|--------|-----------|----------------|------------|
| F      | 0.593     | 0.0203         | 0.66       |
| L      | 0.675     | 0.0431         | 0.86       |
| I      | 0.873     | 0.0669         | 1.60       |
| M      | 0.452     | 0.0378         | 1.95       |
| V      | 0.601     | 0.0615         | 1.52       |
| S(UCN) | 0.415     | 0.0219         | 1.37       |
| P      | 0.501     | 0.0174         | 0.42       |
| T      | 0.555     | 0.0506         | 1.71       |
| A      | 0.462     | 0.0405         | 0.99       |
| Y      | 0.268     | 0.0077         | 0.66       |
| H      | 0.298     | 0.0046         | 0.74       |
| Q      | 0.285     | 0.0058         | 0.74       |
| N      | 0.356     | 0.0259         | 1.10       |
| K      | 0.166     | 0.0072         | 0.66       |
| D      | 0.336     | 0.0029         | 0.59       |
| E      | 0.300     | 0.0068         | 0.37       |
| C      | 0.286     | 0.0033         | 0.71       |
| W      | 0.155     | 0.0013         | 0.13       |
| R      | 0.164     | 0.0015         | 0.21       |
| S(AGY) | 0.338     | 0.0104         | 1.37       |
| G      | 0.243     | 0.0131         | 0.23       |

Table 5

The rank correlation coefficient (with the significance value in parentheses) for each pair of quantities in Table 4

| | |
|---|---|
| Proximity v. Responsiveness | $R = 0.85$ ($p < 10^{-5}$) |
| Proximity v. Mutability | $R = 0.65$ ($p < 0.005$) |
| Responsiveness v. Mutability | $R = 0.74$ ($p < 10^{-4}$) |

Table 6

Ranges of base frequencies at FFD sites in the mammalian genomes: (a) across genome comparison with averages over all plus strand genes; (b) within genome comparison of plus strand genes averaged over all mammalian genomes.

| | (a) Across genome comparison | | (b) Within genome comparison | |
|---|---|---|---|---|
| | Min genome % | Max genome % | Min gene % | Max gene % |
| U | 11.4 | 33.3 | 14.1 | 24.1 |
| C | 13.0 | 43.9 | 23.3 | 32.2 |
| A | 39.0 | 61.5 | 45.2 | 51.8 |
| G | 1.8 | 9.4 | 3.7 | 8.5 |

Figure captions

1. Relationships between the frequencies of bases at fourfold degenerate sites in the plus-strand genes of metazoan mitochondrial genomes. Linear regression lines are shown as a guide to the eye only.

2. Relationship between the frequency of each base at $1^{st}$ and $2^{nd}$ positions and the frequency of the same base at fourfold degenerate sites. Straight lines are predictions using the simple theoretical model in Section 3. The full data set of Metazoa is shown.

3. As figure 2, except that only the Fish data set is shown.

4. The vertebrate mitochondrial genetic code.

5. Principle component analysis of 8 physical properties of amino acids. This demonstrates clear similarity of the groups of amino acids in the first and second columns of the genetic code, and to a lesser extent, the third column also. Fourth column amino acids are very different from one another in physical properties.

6. Relationship between the $1^{st}$ position frequency of U in codons from each of the four columns of the genetic code and the frequency of U at FFD sites in the fish data set. (Note that although only the graphs for U are shown, all four base frequencies are considered when fitting the model).

7. Relationship of the frequencies of serine (UCN codons only), threonine, and alanine to the frequency of U at fourfold degenerate sites in the fish data set. Solid lines are linear regressions.

8. Slopes of the linear regression lines of each amino acid frequency against each base frequency. Black bars – Fish; White bars - Mammals. The amino acids are listed according to their position in the genetic code from column 1 to column 4.

9. Correlation between the proximity and the responsiveness of amino acids. Proximity is a measure of the similarity of an amino acid to its neighbouring amino acids in the genetic code structure. Responsiveness is a measure of the degree to which amino acid frequency varies in response to directional mutation pressure at the DNA level.

10. Comparison of real and simulated data for the Metazoa and Fish data sets. Each graph shows the frequency of U at positions 1 and 2 against U at FFD sites for either a real or a simulated data set. (Note that although only the graphs for U are shown, all four base frequencies are considered when fitting the model).

11. F(X) is the probability that the deviation between a data point and the theory is $\geq$ X. Each graph shows a chi-squared distribution with 3 degrees of freedom, a distribution from a simulated data set (which falls close to the chi-squared distribution), and the distribution for the real data set (which differs significantly from the chi-squared distribution).

12. Frequencies of bases in individual genes averaged over 109 mammalian genomes. Genes on the plus strand are ranked 1-12 in order of increasing $D_{ssH}$, following Reyes *et al.* (1998): COI, COII, ATP8, ATP6, COIII, ND3, ND4L, ND4, ND1, ND5, ND2, CytB. These are shown as open symbols. For comparison, the ND6 gene on the minus strand is shown with filled symbols at gene rank 13. Circles, FFD sites; Squares, $1^{st}$ position; Triangles, $2^{nd}$ position. Linear regression lines through the points for the FFD sites are also shown.

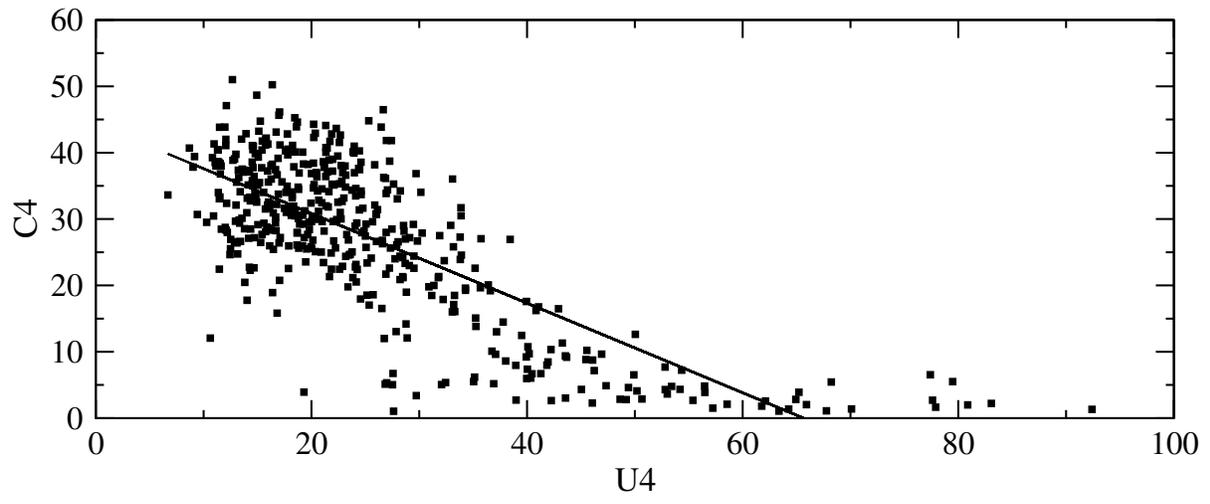

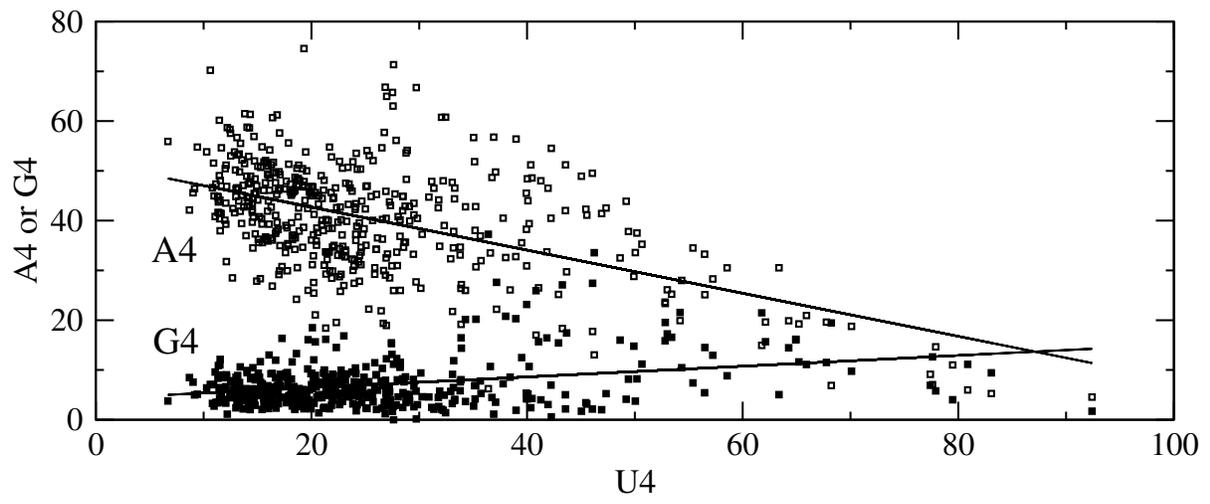

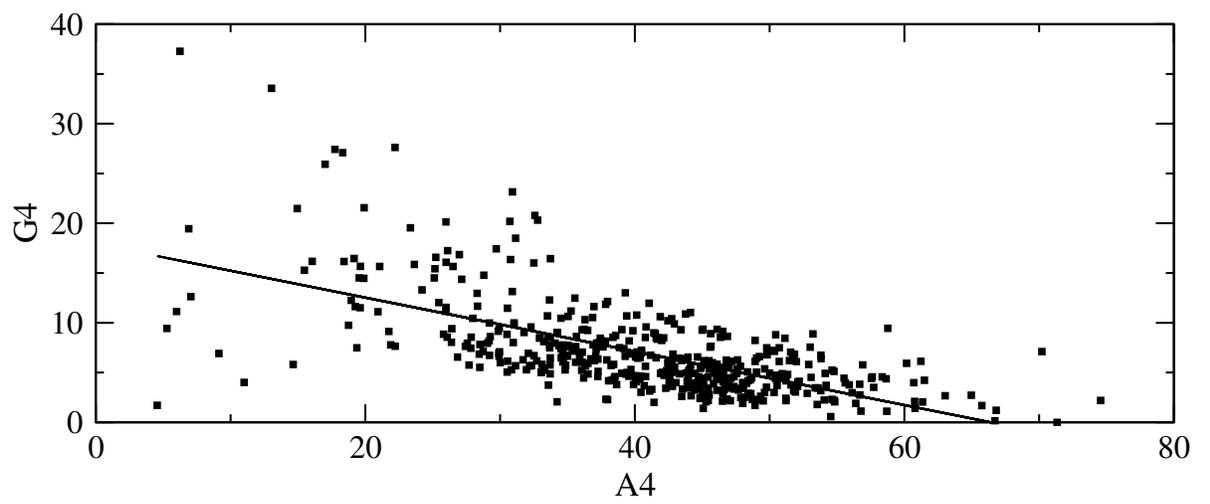

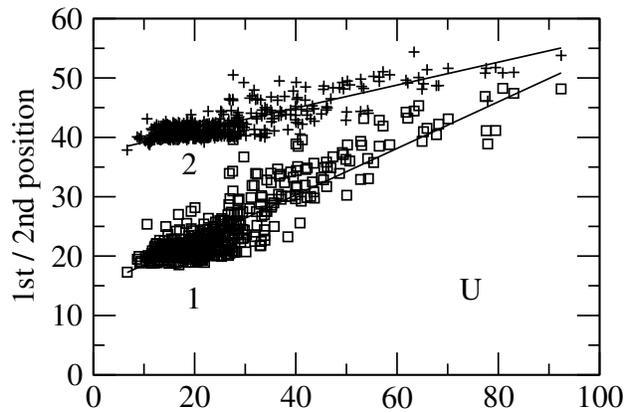
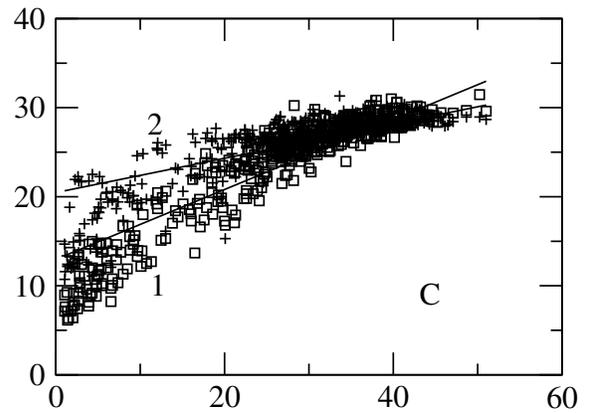
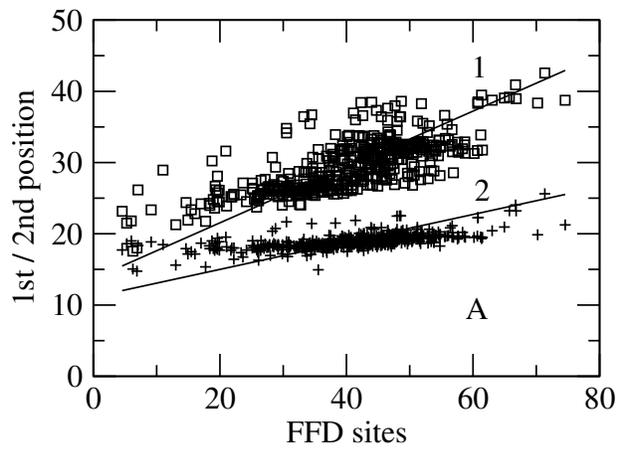
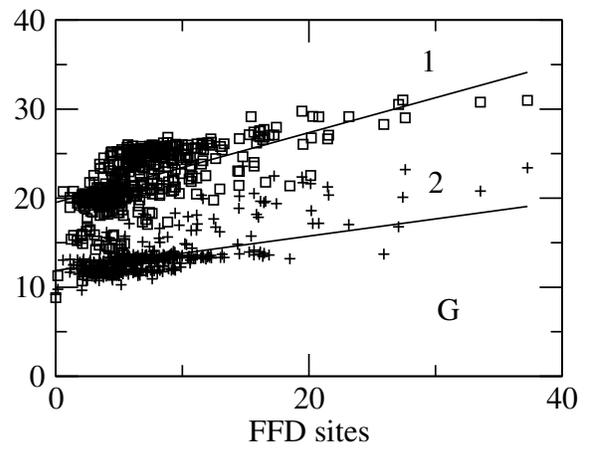

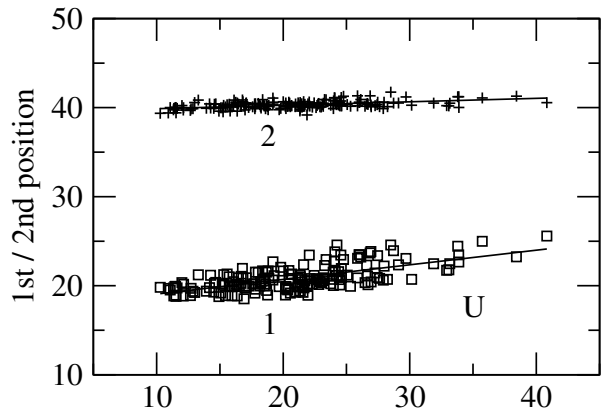
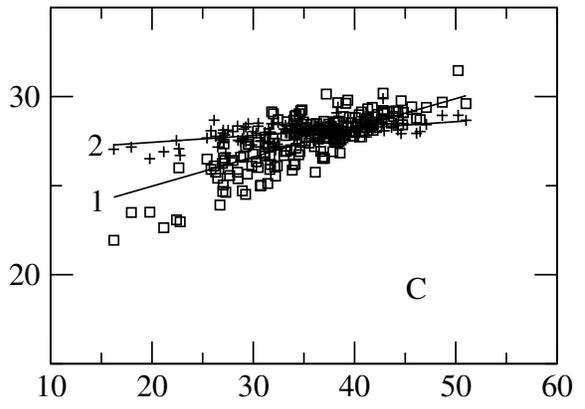
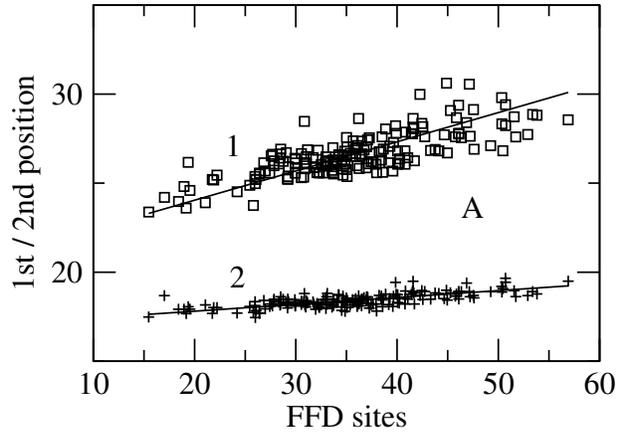
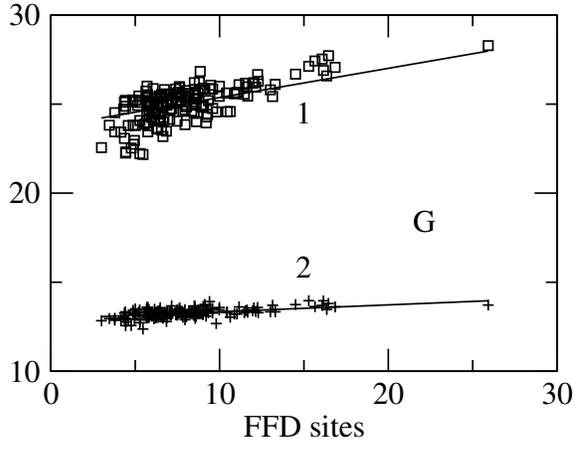

Fig 4

| | | Second Position | | | | |
|---|---|---|---|---|---|---|
| | | U | C | A | G | Third Pos. |
| F | U | F | S | Y | C | U |
| i | | F | S | Y | C | C |
| r | | L | S | Stop | W | A |
| s | | L | S | Stop | W | G |
| t | C | L | P | H | R | U |
| | | L | P | H | R | C |
| P | | L | P | Q | R | A |
| o | | L | P | Q | R | G |
| s | A | I | T | N | S | U |
| i | | I | T | N | S | C |
| t | | M | T | K | Stop | A |
| i | | M | T | K | Stop | G |
| o | G | V | A | D | G | U |
| n | | V | A | D | G | C |
| | | V | A | E | G | A |
| | | V | A | E | G | G |

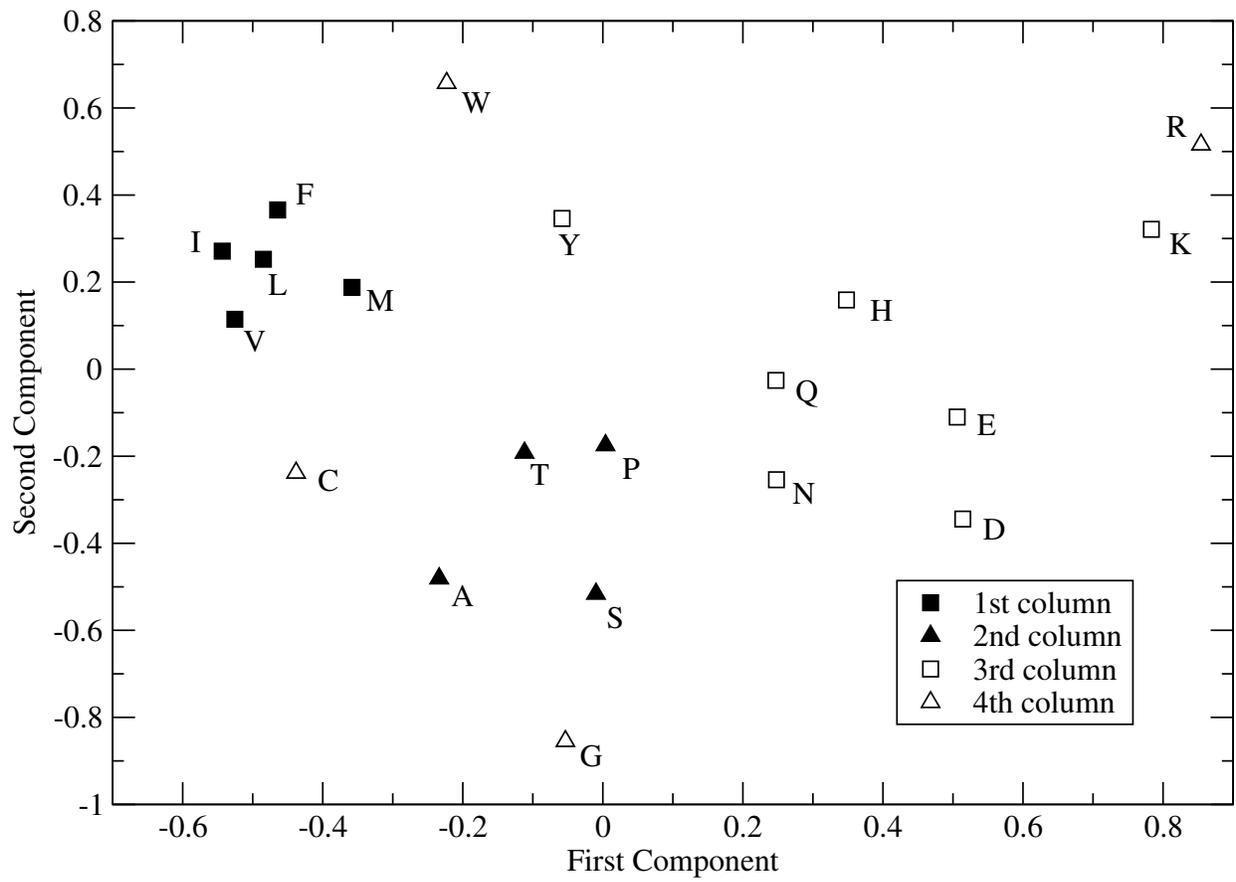

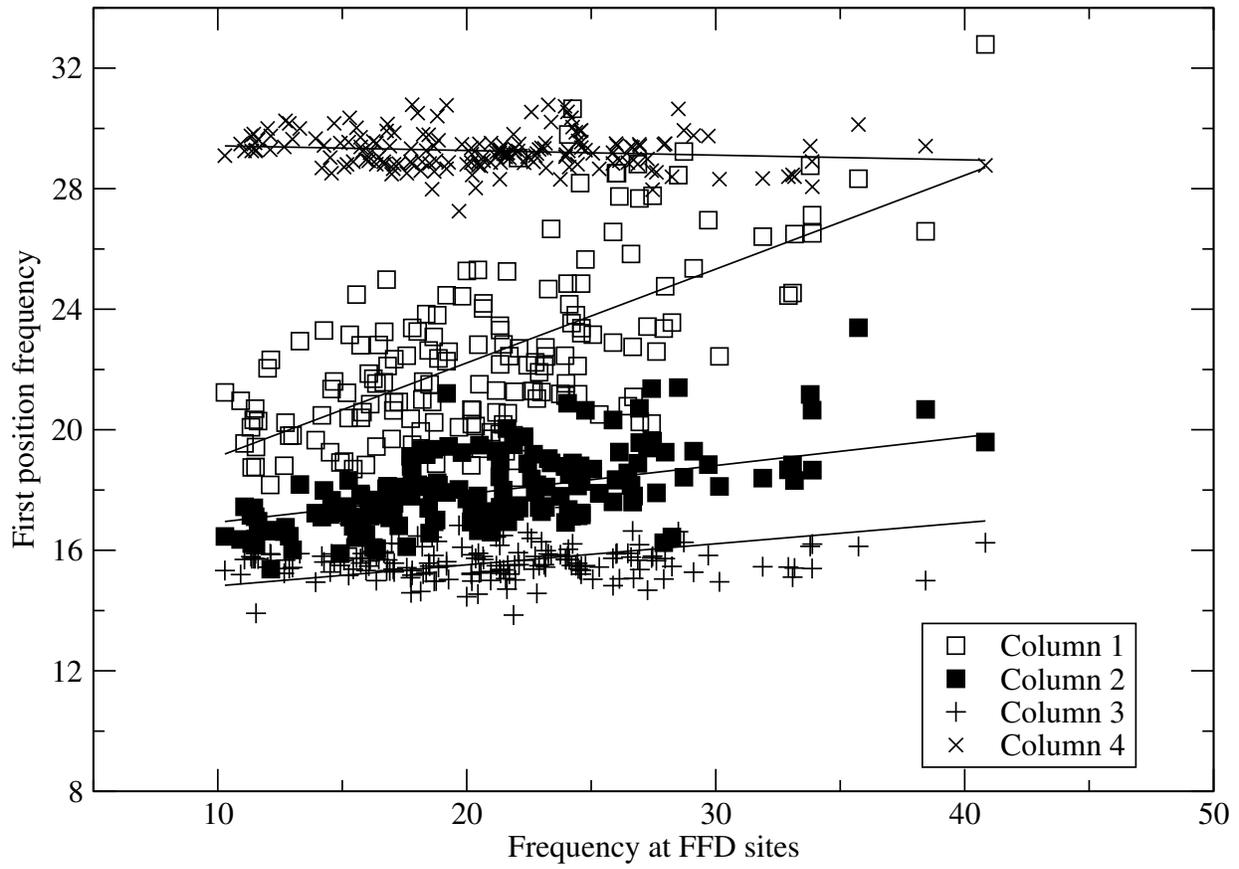

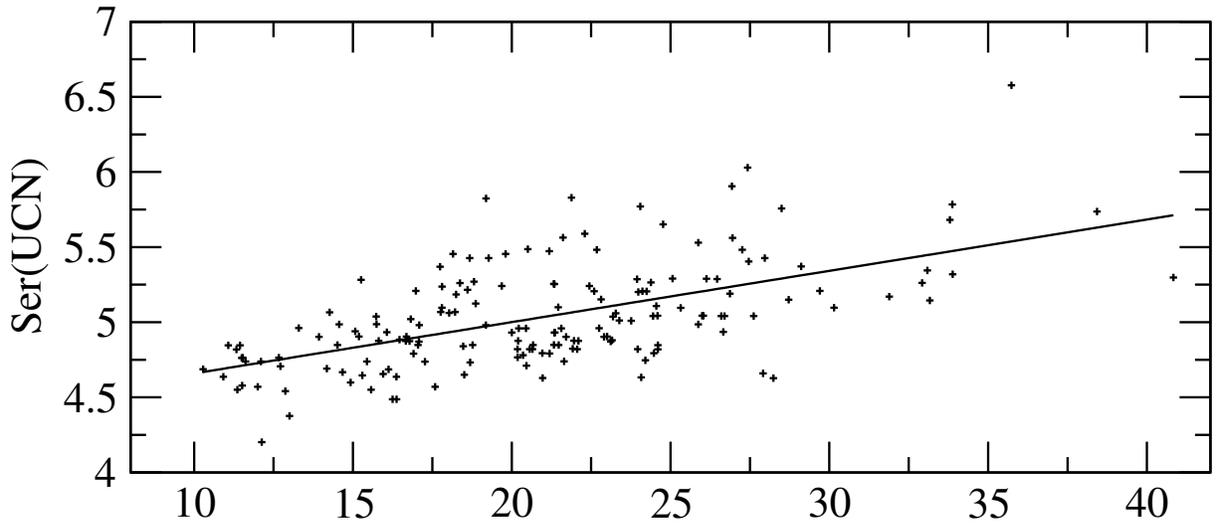
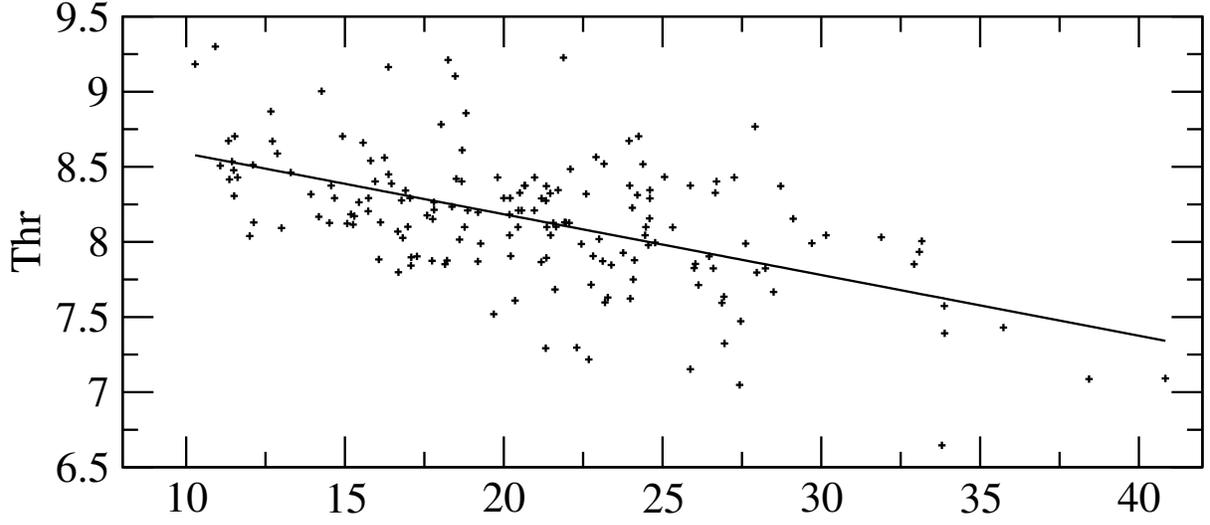
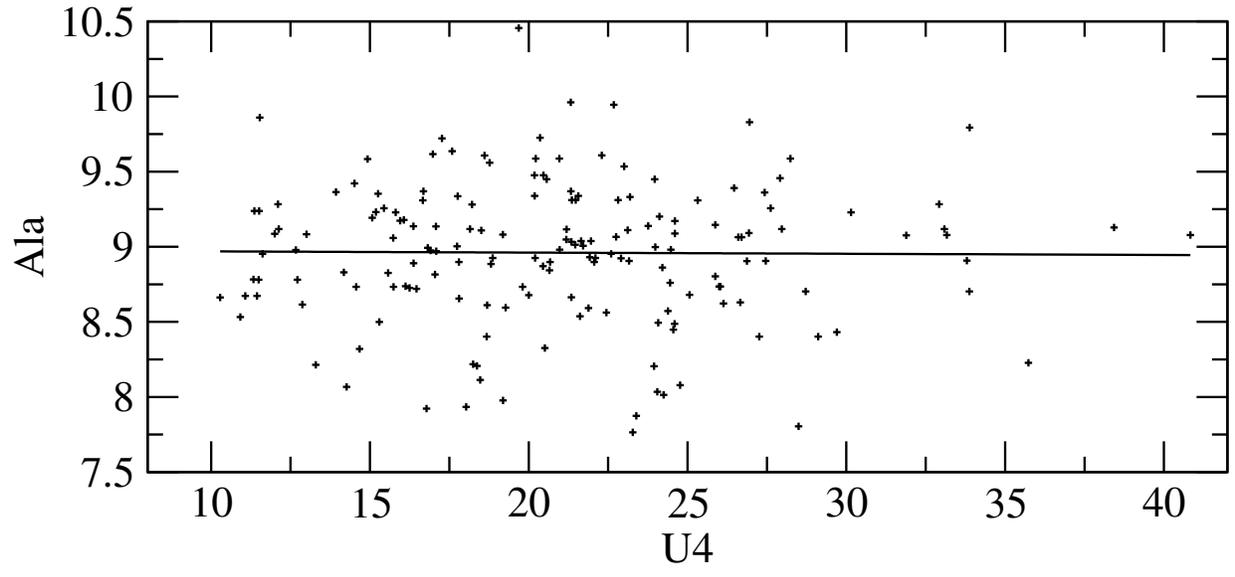

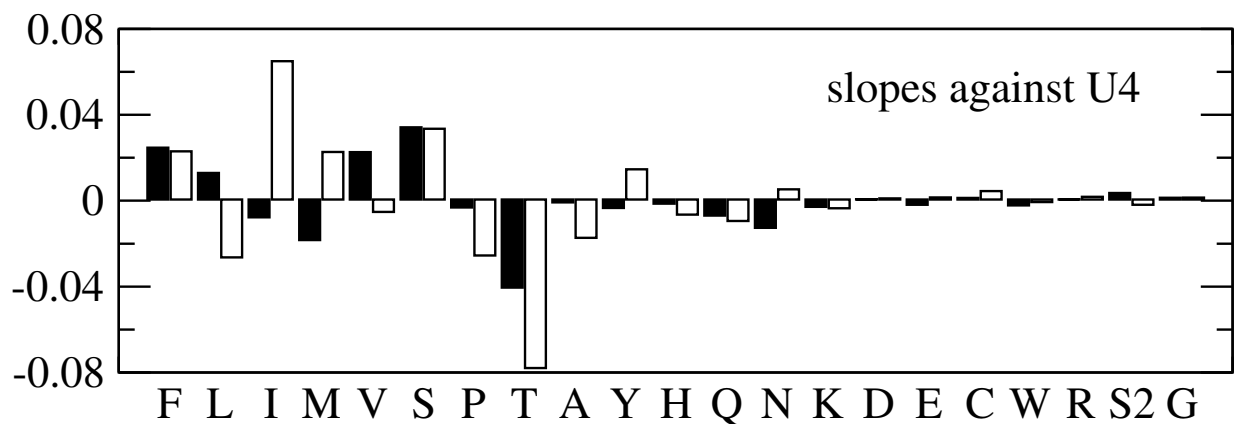

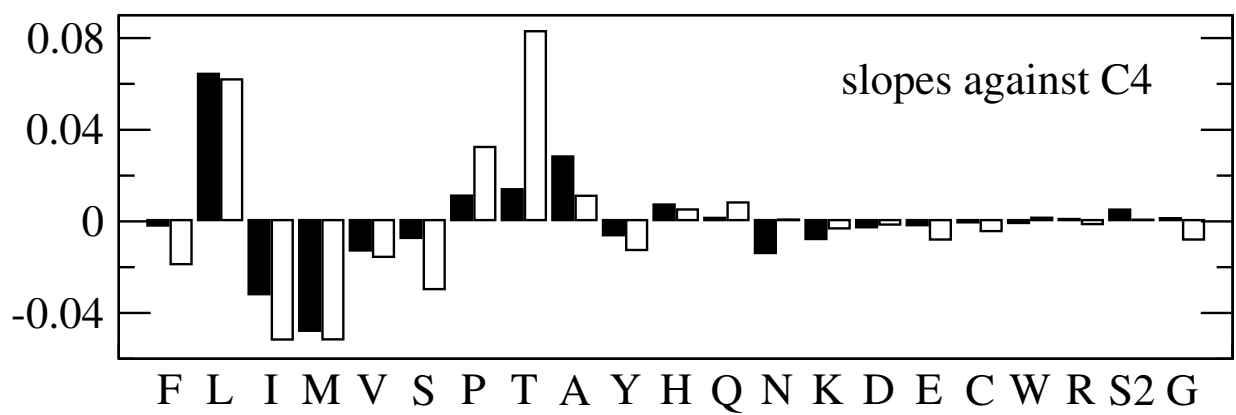

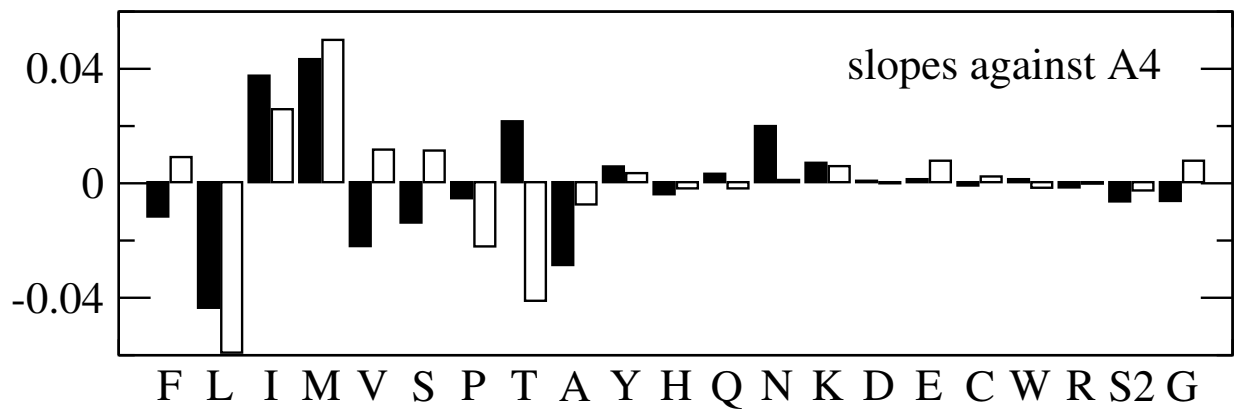

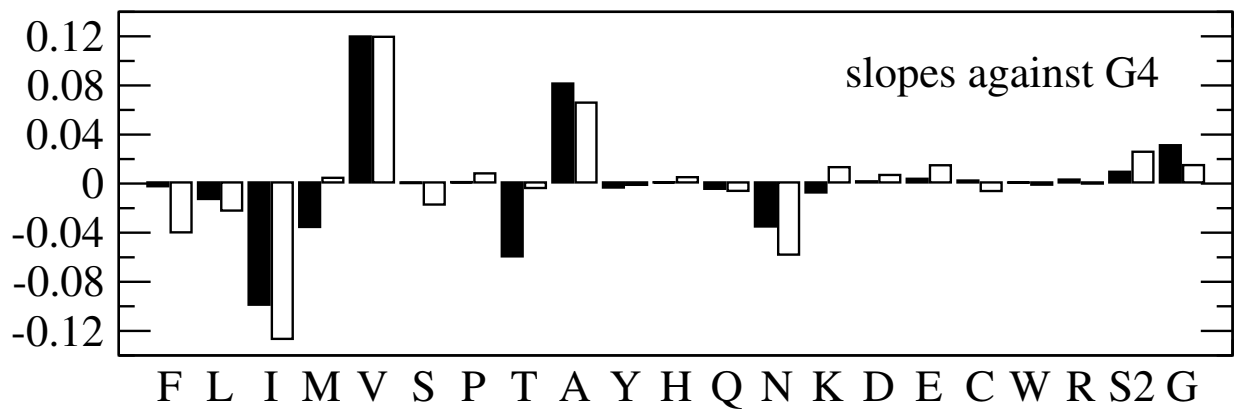

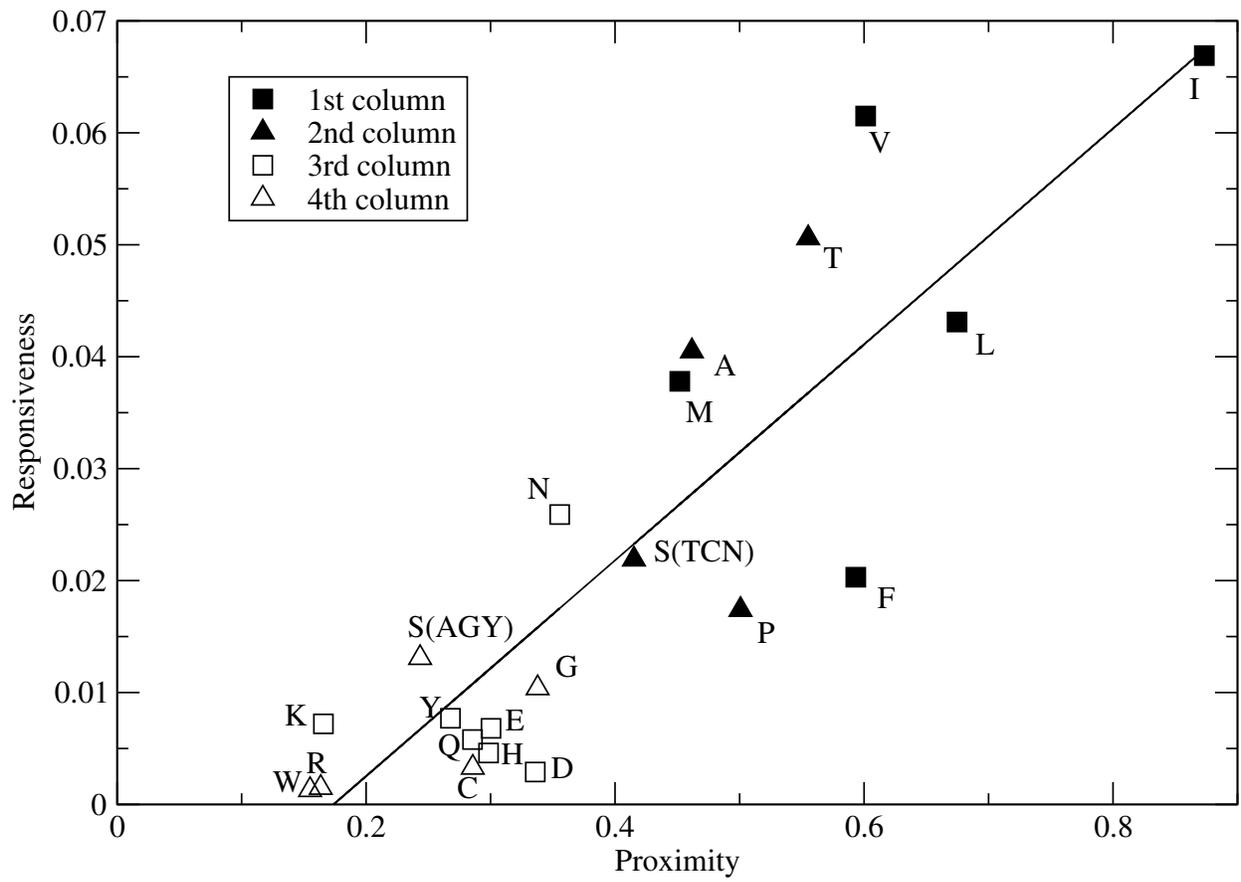

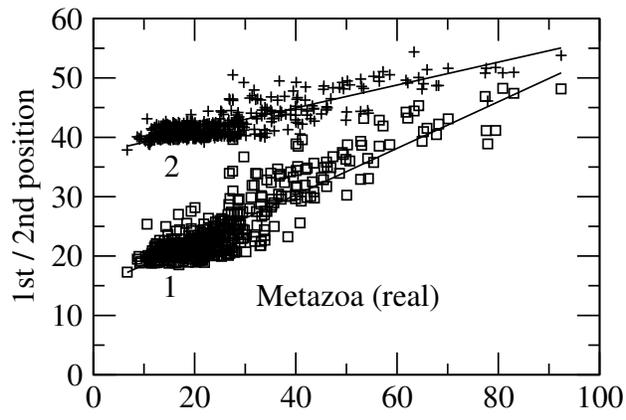
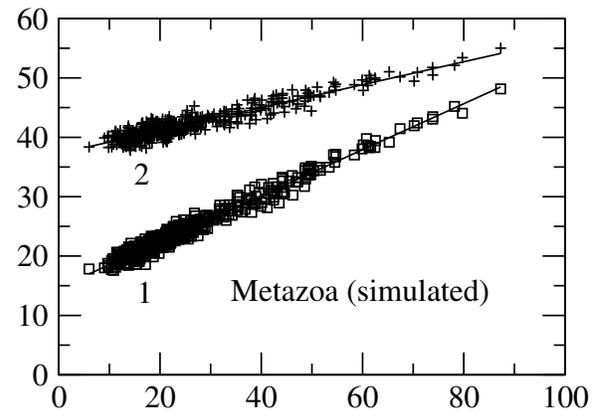
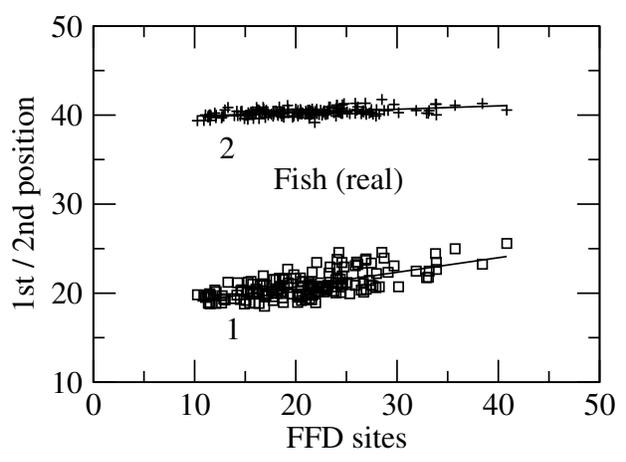
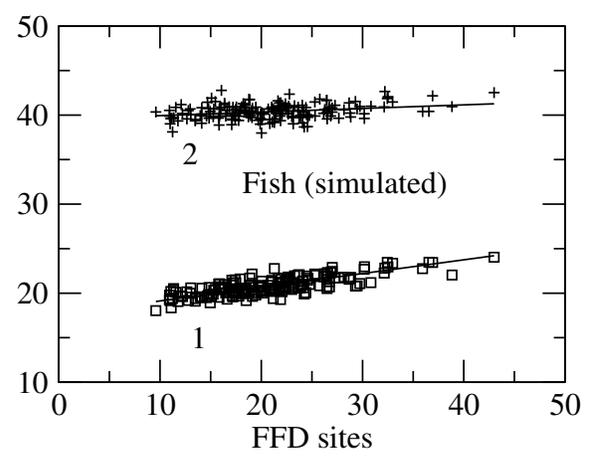

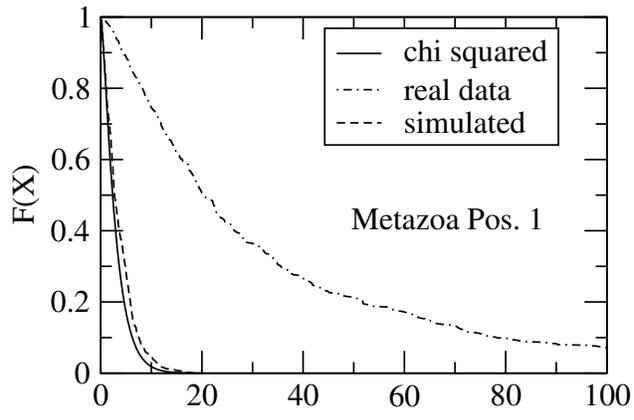
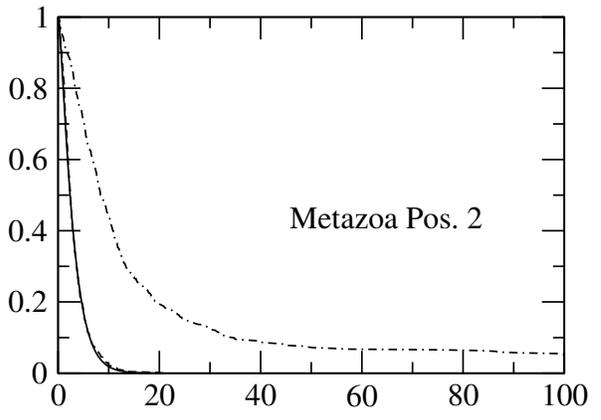
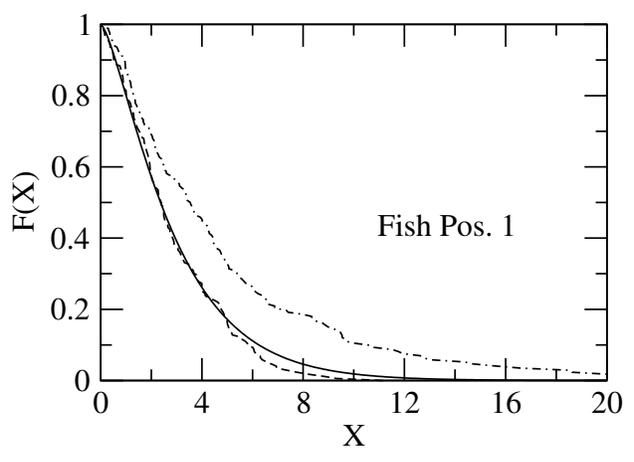
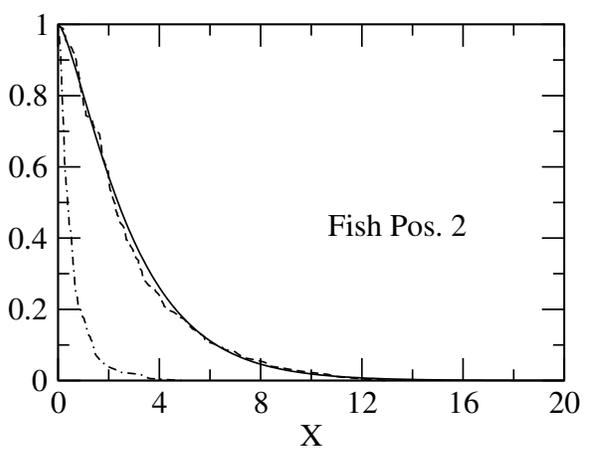

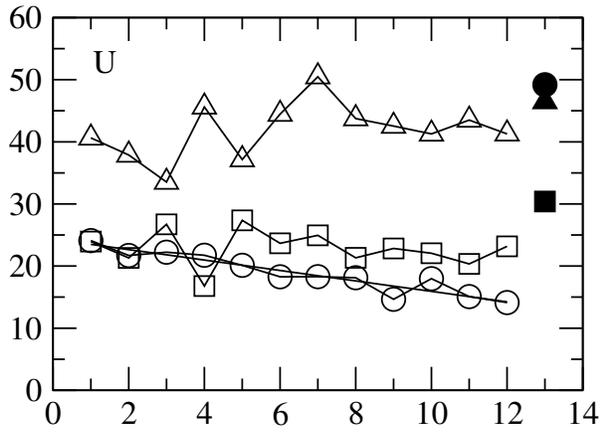

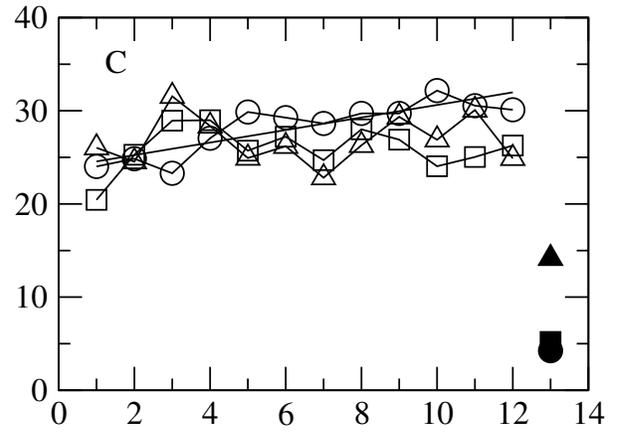

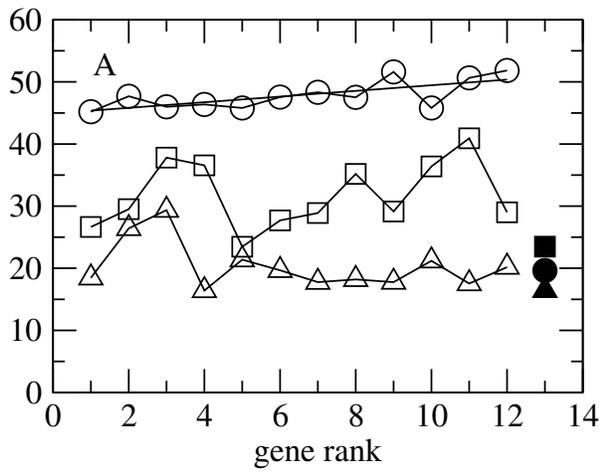

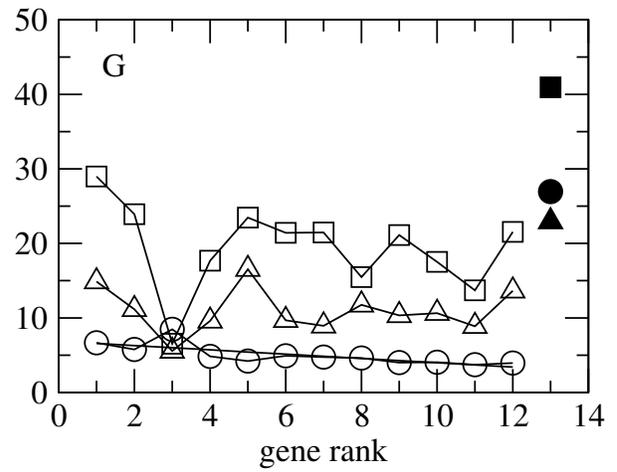